\documentclass [12pt,reqno]{amsart}
\usepackage [dvips] {graphicx}
\usepackage{setspace}
\usepackage{url}
\usepackage{rotating}
\usepackage{multirow}
\usepackage[numbers,compress]{natbib}
\usepackage[gen]{eurosym}
\usepackage{longtable}
\usepackage{pdflscape}
\usepackage{color}
\usepackage{mathtools}
\usepackage{pgfplots}
\usepackage{graphicx}
\usepackage{comment}
\usepackage{caption}
\usepackage{subcaption}
\usepackage{booktabs}
\usepackage{caption}
\usepackage[justification=centering]{caption}
\usepackage{amsaddr}
\usepackage{amssymb}
\usepackage{tikz}
\newcommand*{\tikzhl}[1]{\tikz[baseline=(X.base)] \node[fill=lightgray] (X) {#1};}
\usepackage{pdfpages}
\usepackage{MnSymbol,wasysym}
\usepackage{fontawesome}
\usepackage[top=0.7in, left=1.0in, right=0.7in, bottom=1in]{geometry}
\usepackage{algorithm}
\usepackage{algpseudocode}
\usepackage[toc]{appendix}
\numberwithin{equation}{section}

\newcommand{\Keywords}[1]{\par\noindent
{\small{\em Keywords\/}: #1}}
\begin{document}
\title[\textbf{ZINB}]
 {\textbf{Exponentially Weighted Moving Average Chart using Zero-Inflated Negative Binomial Distribution}}

\author{\textbf{A\lowercase{li} A\lowercase{bbas}$^1$, S\lowercase{ajid} A\lowercase{li}$^{2,*}$, \lowercase{and} I\lowercase{smail} S\lowercase{hah}$^{3,4}$}}

\address{\small{$^{1,2,4}$Department of Statistics, Quaid-i-Azam University, Islamabad 45320, Pakistan\\
		$^3$Department of Statistical Sciences, University of Padua, 35121, Padova, Italy\\
$^1$aliabbas@stat.qau.edu.pk,
$^2$sajidali.qau@hotmail.com, $^3$ismail.shah@unipd.it, $^4$ishah@qau.edu.pk\\ 
*Corresponding Author
}}

\dedicatory{}

\commby{Ali}


\begin{abstract}
Zero-inflated models are frequently used to deal with data having many zeros. A commonly used model for over-dispersed data containing zeros is known as the zero-inflated Poisson model. However, to account for the heterogeneity of counts that leads to excess variance besides inflation of zeros in the data using a more flexible model than the zero-inflated Poisson model, a zero-inflated negative binomial (ZINB) is suggested. In the present study, Shewhart and exponentially weighted moving average (EWMA) control charts are suggested to monitor the ZINB data. The charts are compared using the average run length and standard deviation of run length by using extensive Monte Carlo simulations. Besides a comprehensive simulation study assuming different settings of parameters of ZINB, a real data set is used to show the practicality of the proposed charts. The results indicate that the EWMA chart is better than the Shewhart chart.
\Keywords{Average run length; Exponentially weighted moving average; Shewhart chart; Standard deviation of run length; Zero-inflated negative Binomial}\\
AMS 2010 Classification: 62P30.
\end{abstract}
\maketitle
\section{Introduction}
Statistical process control (SPC) uses statistical methods to monitor, control, and improve the quality of a process. A control chart is a tool of SPC that is used for monitoring process changes or variations. The characteristics to be monitored of a process and how these samples are taken both influence the choice of the control charts. Shewhart ‘p’ and ‘np’ control charts are frequently used to monitor the processes under the assumption of binomial distribution. These charts only take the latest observations in a series of points, and they fail to detect small and moderate shifts in the process. 

In real-life situations, we often encounter data having excessive zeros and several statistical models can be used for modeling such data sets. However, the non-conforming fraction is quite low in many high-yield processes because of technological improvement. The excess of zeros causes an over-dispersion, which undermines the zero-inflated Poisson (ZIP) application. Also, the control charts based on the binomial distribution have tighter control limits that result in a higher false alarm rate. Therefore, these charts cannot be applied effectively. To account for the inflation of zeros besides over-dispersion, zero-inflated negative binomial (ZINB) model is more appropriate than the simple Poisson and negative binomial distributions. 
	Saghir and Lin\cite{New8} and Mahmood and Xie\cite{New1} presented a comprehensive overview of control charts for dispersed count data. Chananet et al.\cite{Chananet} proposed EWMA chart assuming ZNIB and compared Markov Chain approach with Monte Carlo simulations for the ARL computation. However, they did not assess the role of smoothing parameter on the ARL. Katemee and Mayureesawan\cite{New5} discussed Shewhart-type charts assuming ZIP process. Fatahi et al.\cite{New3} discussed an exponentially weighted moving average (EWMA) chart for the ZIP random variable to monitor rare health-related events. Rakitzis and Castagliola\cite{New9} discussed the effect of estimation on Shewhart-type control charts for zero-inflated processes. Mukherjee and Rakitzis\cite{New2} proposed progressive monitoring schemes using the Mahalanobis distance between the true value of parameter and its maximum likelihood estimate for both parameters of the ZIP. They also proposed two separate likelihood ratio tests for post-signal follow-up. 
	 Urbieta et al.\cite{New7} compared EWMA and cumulative sum (CUSUM) charts using the generalized linear model with negative binomial distribution for the number of hospitalizations. 
	 Lai et al.\cite{New10} proposed Pearson and deviance residuals based EWMA charts with risk-adjustments for the ZIP process. 
	 Hu and Liu\cite{New13} studied an upper-sided EWMA control chart using a weighted score test statistic. The proposed control chart can detect the positive shifts in parameters of the Poisson part in ZIP model using covariates. Asghar et al.\cite{New14} used the Poisson hurdle model to study GLM-based Shewhart-type control charts. 

Greene\cite{greene1994accounting} noticed the suitability of the ZINB model for dealing with excess zeros and over-dispersion. The ZINB distribution is made up of a Poisson gamma mixed distribution that contains a dispersion parameter $k$. The ZINB distribution is reduced to the ZIP distribution as $k\rightarrow0$. Sharma and Landge\cite{sharma2013zero} used the ZINB model for analyzing traffic accidents. 
Jansakul and Hinde\cite{jansakul2008score} proposed a generalized score test statistic for comparing the ZINB regression model to the NB model. Zulkifli et al.\cite{zulkifli2011zero} compared the ZIP and ZINB using the theft insurance data. Aa and Naing\cite{aa2012analysis} compared the mortality rates using negative binomial (NB) and ZINB models. Saputro and Qudratullah\cite{saputro2021estimation} discussed estimation techniques for the ZINB regression. For more details, interested readers are referred to Saputro and Qudratullah\cite{cheung2002zero}, Rose et al.\cite{rose2006use}, Hall\cite{hall2000zero}, Yau and Lee\cite{yau2001zero}, Hall and Wang\cite{hall2005two}, Williamson et al.\cite{williamson2007power}, Hashim et al.\cite{hashim2021application}, Adarabioyo and Ipinyomi\cite{adarabioyo2019comparing}, Sheu et al.\cite{sheu2004effect}, and references cited therein.


The main objective of the study is to propose Phase-II Shewhart and EWMA control charts for the ZINB distribution and compare their performance by using the average run length (ARL). To the best of our knowledge, no such comprehensive study for the monitoring of ZINB process with different smoothing parameters exists in the literature to date. The rest of the article is organized as follows. In Section 2, we present the construction of EWMA and Shewhart charts using the ZINB. The result based on an extensive simulation study are discussed in Section 3. A real data example is discussed in Section 4. The study is concluded in Section 5. 
 \section{The ZINB distribution and Control Charts}
The ZINB distribution is a mixture of a distribution degenerated at zero and an ordinary negative binomial distribution. The degenerate distribution captures the excessive number of zeros, which exceeded those predicted by the negative binomial distribution. The ZINB  model assumes that there are two distinct data generation processes. The result of a Bernoulli trial is used to determine which of the two processes is used. For $i$th observation, a zero count is generated with probability $\theta_i$, and with a probability of $(1-\theta_i)$, the response is generated from the negative binomial with success probability $p$ of the successive trials. It is worth mentioning that the zero counts are generated from both processes, and a probability is estimated for whether zero counts are from the first or the second process. Let $Y_i$ be a random variable that follows the ZINB distribution. Then, the probability mass function of $Y_i$ is given by \citep{New17} 
\begin{gather}
Pr(Y_i=0)=\theta+(1-\theta)p^k\\
P(Y_i=y)=(1-\theta)\binom{y+k-1}{y}p^k(1-p)^y, y=1,2,3,\cdots
\end{gather}
The mean and variance of $Y_i$ are $E(Y_i)=k(1-\theta)(1-p)/p$ and $Var(Y_i)=k(1-\theta)(1-p)[1+(1-p)\theta k]/p^2$, where $p$ is probability of success in each trial, $k$ is the over-dispersion parameter, and $\theta$ is the probability of extra zeros. The ZINB distribution reduces to the ZIP distribution as $k\rightarrow0$. Also, for $\theta=0$, the ZINB reduces to the ordinary negative binomial distribution, where $Y$ is the number of failures until the occurrence of $k$ successes with $Y\in \{0,1,2,\cdots\}$.
\subsection{The ZINB-EWMA control chart}    
Let $Y_i$ follow the ZINB$(p, k, \theta)$ for $i=1,2,3,\cdots$ with mean $E(Y_i)=k(1-\theta)(1-p)/p$, then the ZINB-EWMA statistic for sample mean is defined as
\begin{equation}\label{chart}
Z_i=\lambda \bar Y_i+(1-\lambda)Z_{i-1}                   
\end{equation}
where $0< \lambda<1$ is the smoothing factor, $Z_i$ denote the EWMA statistic at time $i$, and $Z_{i-1}$ denote the EWMA statistic at time $i-1$. The in-control (IC) expected value and variance of the statistic $Z_i$ are $E(Z_i|IC) =E(Y_i)=k(1-\theta)(1-p)/p$ and $Var(Z_i|IC) =  \dfrac{\lambda}{n(2-\lambda)}[k(1-\theta)(1-p)[1+(1-p)\theta k]/p^2]$. Hence, the upper control limit (UCL), lower control limit (LCL), and  the centre line (CL) of the ZINB-EWMA chart are given by
\begin{align}
&UCL=\frac{k(1-\theta)(1-p)}{p}+L\sqrt{\frac{\lambda k(1-\theta)(1-p)[1+(1-p)\theta k]}{np^2(2-\lambda)}}\label{limit}\\
&CL = \frac{k(1-\theta)(1-p)}{p}\label{limit1}\\
&LCL=\max\left\{0,\frac{k(1-\theta)(1-p)}{p}-L\sqrt{\frac{\lambda k(1-\theta)(1-p)[1+(1-p)\theta k]}{np^2(2-\lambda)}}\right\}\label{limit2}
\end{align}
It is worth mentioning that we use only the UCL as the EWMA statistic is always positive. The above limits are applicable if $n>1$ and these limits reduce to a singe observation monitoring assuming $n=1$. To obtain the Shewhart-ZINB chart, substitute $\lambda=1$ in Eqs. \ref{chart} and \ref{limit}.
\section{Performance of the ZINB-EWMA and ZINB-Shewhart control charts}
This section investigates the run-length properties of the ZINB-EWMA and ZINB-Shewhart charts. In particular, we study the IC and out-of-control (OOC) ARL and standard deviation of the run length (SDRL).  

The most widely used performance measure of a control chart is the ARL, which is the average of the number of samples taken before an OOC signal is detected by a chart. When a process is IC, $ARL_0$ is desired to be large, which would lead to fewer false signals. However, if a process is OOC, then $ARL_1$ is desired to be relatively small. It is important to set a common $ARL _0$ for two control charts to compare their performance. A chart with a lower $ARL_1$ value for a specific shift is more effective in terms of shift detection than the other competing charts.
\subsection{Algorithm for run-length distribution}
This study uses the Monte Carlo simulation method to calculate the ARL and SDRL of the ZINB-EWMA and ZINB-Shewhart charts. The simulation algorithm consists of the following steps. 
\begin{enumerate}
	\item Fix $k,p,\theta$ and generate 10,000 random samples ${\bf Y}_i, i = 1, 2, \cdots, 10,000$ from a ZINB $(k, p, \theta)$ distribution. 
	\item For a given $\lambda$ and the appropriate value of $L$ to achieve the desired $ARL_0$ of the ZINB-EWMA, the control limits are calculated using Eq. \ref{limit}. Then, the monitoring statistic is compared with the corresponding values of the UCL. 
	\item To compute the run length, record the sample number at which the first-time monitoring statistic falls outside the UCL.
	\item After running 10,000 iterations of steps 1-3, take the average of run length to obtain $ARL_0$  and compare it to the desired value. 
	\item If the desired $ARL_0$ is not achieved, change the value of $L$ and repeat steps 2-4. Otherwise, use this UCL to monitor OOC data. 
	\item To compute $ARL_1$ and $SDRL_1$, generate data from ZINB $(k_1, p_1, \theta_1)$, where $k_1, p_1, \theta_1$ denotes OOC parameters, and record the first OCC signal by plotting monitoring statistic against the UCL. Repeat this step 10,000 times to obtain $ARL_1$ and $SDRL_1$.
\end{enumerate}
It is worth mentioning that 10,000 replications are sufficient to compute ARL and SDRL, although Mundform et al.\cite{mundform2011number} noticed that 5000 replications are sufficient to achieve the desired $ARL_0$ with a low error level.
\subsection{Results} 
In this section, the behavior of ARLs and SDRLs of the ZINB-EWMA and ZINB-Shewhart is discussed for different combinations of parameters $(p ,k, \theta)$, using $n=1$ for $ARL_0 = 500$. 

Table \ref{tab:1} tabulates the ARL and SDRL of the ZINB-EWMA and ZINB-Shewhart charts for the IC and OOC processes. Assuming (n, k, $p_0$, $\theta$)=(1, 1, 0.40, 0.85), $\lambda$=0.05, and L=3.105, we have the $ARL_0$=500.81 with $SDRL_0$=501.87 for the ZINB-EWMA chart. However, (n, k, $p_0$, $\theta$)=(1, 1, 0.40, 0.85) with $\lambda$=0.15 and L=4.733, we obtained $ARL_0$= 500.94 with $SDRL_0$=503.13, respectively. For the ZINB-Shewhart chart with the aforementioned specifications and $\lambda$=1, the L is 5.067. However, $ARL_0$=396.01 with $SDRL_0$=394.49, is achieved for the IC parameters. Therefore, it is noticed that IC ARL is not achieved in the case of the ZINB-Shewhart chart.

For the OOC case with $p_1$=0.38, the $ARL_1$ and $SDRL_1$ for the ZINB-EWMA chart with $\lambda$=0.05 and L=3.015 are 360.24 and 356.12, respectively. Assuming (n, k, $p_0$, $\theta$)=(1, 1, 0.40, 0.85) and $p_1$=0.38, the $ARL_1$ and $SDRL_1$ for the ZINB-EWMA chart with $\lambda$=0.15 and L=4.733 are 370.51 and 375.75, respectively. Similarly, we calculated the $ARL_1$ and $SDRL_1$ of the ZINB-EWMA for the different values of $\lambda$ like (0.10,0.20,0.25,0.50,0.80) and it is noticed that a small smoothing parameter value performs better than a large value.  For the ZINB-Shewhart chart with $p_1$=0.38, $\lambda$=1, and L=5.067, the $ARL_1$ and $SDRL_1$ are 302.58 and 306.35, respectively. Likewise, assuming $p_1$=0.35 the $ARL_1$ and $SDRL_1$ of the ZINB-Shewhart with $\lambda$=1 and L=5.067 are 208.52 and 209.84, respectively. Although the OOC ARL of the ZINB-Shewhart is smaller than the ZINB-EWMA chart, the ZINB-Shewhart is not preferred because it does not achieve the desired $ARL_0= 500$.

Similarly, Tables \ref{tab:2}-\ref{tab:3} list the performance of the charts assuming $k=2$ and 5, respectively. In particular, from Table \ref*{tab:2} with $\theta$)=(1, 2, 0.40, 0.85), $\lambda$=0.10, and L=3.742, the $ARL_0$=499.93 with $SDRL_0$=500.31 is calculated. However, with $\lambda$=1 and L=7.355 for the ZINB-Shewhart, the $ARL_0$ is 521.21 with $SDRL_0$=529.18. Assuming $p_1$=0.38 for the OOC case, the $ARL_1$ and $SDRL_1$ for the ZINB-EWMA chart with $\lambda$=0.10 and L=3.742 are 338.58 and 338.07, respectively. For the ZINB-Shewhart chart with $\lambda$=1 and L=7.335, we obtained 366.24 and 368.53 as the $ARL_1$ and $SDRL_1$. Again, it is noticed that the ZINB-EWMA performs well as compared to the ZINB-Shewhart. Furthermore, the values of $L$ have a decreasing pattern as we increased the value of $k$.

Next, Tables \ref{tab:7}, \ref{tab:8}, and \ref{tab:9} list the ARL and SDRL results of the ZINB-EWNA and the ZINB-Shewhart for n=1, k =(1,2,5) and $\theta$=0.80. For example, Table \ref{tab:7} with $\lambda$=0.20 and L=4.980 gives the $ARL_0$=500.25 with $SDRL_0$=497.61. Also, with $\lambda$=1 and L=7.995 for the ZINB-Shewhart, we have the $ARL_0$=499.64 with $SDRL_0$=494.61. For the OOC case $p_1=0.38$, the $ARL_1$ and $SDRL_1$ for the ZINB-EWMA chart with $\lambda$=0.20 and L=4.980 are 363.91 and 360.72, respectively. For the ZINB-Shewhart chart with $\lambda$=1 and L=7.995, we have $ARL_1=372.28$ and $SDRL_1=376.29$. It is worth mentioning that although the $ARL_0=500$ for the Shewhart chart, the $ARL_1$ values of the EWMA chart are smaller than the counterpart. It is also noticed that for $n=1$ and $k=2$, and 5, the ZINB-Shewhart chart does not achieve the desire ARL and generally, the ZINB-EWMA performs well as compared to the ZINB-Shewhart. 

Tables \ref{tab:13} to \ref{tab:15} show the ARL values for the ZINB-EWMA and ZINB-Shewhart charts for n= 1, k=1,2,5 and $\lambda$ =0.75. For example, consider Table \ref{tab:14} where the IC process assumes (n, k, $p_0$, $\theta$)=(1, 2, 0.40, 0.75). Assuming $\lambda$=0.10 and L=3.465 gives an $ARL_0$=500.92 with $SDRL_0$=498.38. However, with $\lambda$=1 and L=6.295, the ZINB-Shewhart produces $ARL_0$=497.27 with $SDRL_0$=498.81. Next, considering $p_1$=0.38 as the OOC case, the $ARL_1$ and $SDRL_1$ for the ZINB-EWMA chart with $\lambda$=0.10 and L=3.465 are 314.31 and 309.56, respectively. However, the ZINB-Shewhart chart assuming $\lambda$=1 and L=6.295 produces $ARL_1=336.73$ and $SDRL_1=342.13$. 

From the above tables, it is clear that the ZINB-EWMA chart outperforms the ZINB-Shewhart chart. The performance of the ZINB-EWMA deteriorates as the value of the smoothing parameter increased. This is further supplemented by the value of $L$, which increases as $\lambda$ increased.
	\begin{table}[H]
	\centering
	\scriptsize
	\caption{Performance of ZINB EWMA and Shewhart charts for k=1, n=1, and $\theta$=0.85}
	\begin{tabular}{cccccccccc}
		\toprule
		L   &     & 8.435 & 3.105 & 4.037 & 4.733 & 5.265 & 5.814 & 7.153 & 8.207 \\
		\midrule
		UCL &     & 7.9988 & 0.6832 & 1.0785 & 1.4670 & 1.8424 & 2.2502 & 4.0310 & 6.4007 \\
		\midrule
		$\lambda$ &     & 1   & 0.05 & 0.10 & 0.15 & 0.20 & 0.25 & 0.50 & 0.80 \\
		\midrule
		p   &     & Shewhart & EWMA & EWMA & EWMA & EWMA & EWMA & EWMA & EWMA \\
		\midrule
		0.40 & ARL & 396.01 & 500.81 & 500.72 & 500.94 & 500.61 & 499.48 & 498.36 & 537.05 \\
		& SDRL & 394.49 & 501.87 & 497.63 & 503.13 & 503.25 & 502.34 & 498.23 & 533.93 \\
		0.38 & ARL & 302.58 & \tikzhl{360.24} & 364.66 & 370.51 & 374.22 & 372.03 & 375.14 & 404.3 \\
		& SDRL & 306.35 & 356.12 & 364.25 & 375.75 & 378.37 & 376.9 & 378.01 & 408.81 \\
		0.35 & ARL & 208.52 & \tikzhl{229.21} & 232.79 & 237.13 & 241.91 & 242.75 & 247.03 & 265.85 \\
		& SDRL & 209.84 & 228.37 & 235.92 & 237.71 & 243.94 & 245.51 & 248.33 & 267.64 \\
		0.33 & ARL & 162.38 & \tikzhl{173.16} & 177.91 & 180.65 & 183.82 & 186.49 & 191.18 & 205.74 \\
		& SDRL & 162.63 & 171.55 & 177.16 & 179.76 & 183.17 & 185.76 & 192.07 & 206.71 \\ \hline
	\end{tabular}%
	\label{tab:1}%
\end{table}%
	\begin{table}[H]
	\centering
	\scriptsize
	\label{tab:2}
	\caption{Performance of ZINB EWMA and Shewhart charts for k=2, n=1, and $\theta$=0.85}
	\begin{tabular}{cccccccccc}
		\toprule
		L   &     & 7.335 & 2.951 & 3.742 & 4.296 & 4.740 & 5.111 & 6.381 & 7.054 \\
		\midrule
		UCL &     & 11.5073 & 1.1623 & 1.7441 & 2.2940 & 2.8318 & 3.3621 & 6.0036 & 9.1324 \\
		\midrule
		$\lambda$ &     & 1   & 0.05 & 0.10 & 0.15 & 0.20 & 0.25 & 0.50 & 0.80 \\
		\midrule
		p   &     & Shewhart & EWMA & EWMA & EWMA & EWMA & EWMA & EWMA & EWMA \\
		\midrule
		0.40 & ARL & 521.21 & 499.76 & 499.93 & 500.05 & 500.20 & 500.98 & 498.99 & 499.63 \\
		& SDRL & 529.18 & 500.31 & 505.07 & 509.56 & 512.55 & 511.62 & 506.41 & 505.27 \\
		0.38 & ARL & 366.24 &\tikzhl{330.20} & 338.58 & 340.45 & 342.21 & 345.49 & 348.18 & 351.11 \\
		& SDRL & 368.53 & 326.8 & 338.07 & 345.29 & 346.57 & 349.37 & 355.41 & 353.55 \\
		0.35 & ARL & 221.24 & \tikzhl{193.63} & 199.99 & 202.57 & 203.56 & 205.92 & 208.57 & 212.47 \\
		& SDRL & 224.41 & 188.63 & 195.81 & 200.98 & 203.41 & 207.22 & 210.41 & 214.78 \\
		0.33 & ARL & 163.21 & \tikzhl{140.79} & 144.46 & 147.35 & 148.63 & 149.83 & 151.07 & 155.21 \\
		& SDRL & 163.73 & 135.84 & 142.17 & 147.70 & 148.97 & 150.65 & 151.42 & 156.02 \\ \hline
	\end{tabular}%
	\label{tab:2}
\end{table}%
 \begin{table}[H]
	\centering
	\scriptsize
	\label{tab:3}
	\caption{Performance of ZINB EWMA and Shewhart charts for n=1, k=5, and $\theta$=0.85}
	\begin{tabular}{cccccccccc}
		\toprule
		L   &     & 5.949 & 2.792 & 3.416 & 3.829 & 4.141 & 4.399 & 5.210 & 5.766 \\
		\midrule
		UCL &     & 19.9226 & 2.5376 & 3.6012 & 4.5701 & 5.4865 & 6.3787 & 10.6296 & 16.0011 \\
		\midrule
		
		$\lambda$ &     & 1   & 0.05 & 0.10 & 0.15 & 0.20 & 0.25 & 0.50 & 0.80 \\
		\midrule
		p   &     & Shewhart & EWMA & EWMA & EWMA & EWMA & EWMA & EWMA & EWMA \\
		\midrule
		0.40 & ARL & 499.02 & 500.17 & 500.14 & 500.77 & 501.57 & 500.19 & 501.53 & 534.04 \\
		& SDRL & 494.64 & 496.61 & 508.17 & 515.01 & 513.81 & 514.06 & 503.51 & 533.97 \\
		0.38 & ARL & 303.39 & \tikzhl{298.43} & 303.71 & 305.44 & 304.76 & 304.76 & 302.29 & 324.83 \\
		& SDRL & 301.45 & 295.77 & 304.99 & 307.46 & 305.58 & 302.91 & 301.25 & 324.05 \\
		0.35 & ARL & 155.79 & 157.59 & 158.71 & 158.75 & 157.92 & 156.85 & \tikzhl{154.38} & 163.06 \\
		& SDRL & 155.63 & 153.31 & 157.23 & 156.95 & 157.45 & 156.54 & 154.35 & 163.27 \\
		0.33 & ARL & 104.02 & 110.66 & 109.31 & 108.81 & 107.23 & 106.41 & \tikzhl{102.99} & 109.07 \\
		& SDRL & 102.19 & 104.91 & 106.81 & 106.11 & 105.86 & 105.01 & 100.37 & 106.83 \\ \hline
	\end{tabular}%
	\label{tab:3}
\end{table}%

\begin{table}[H]
	\centering
	\scriptsize
	\label{tab:7}
	\caption{Performance of ZINB EWMA and Shewhart charts for n=1, k=1, and $\theta$=0.80}
	\begin{tabular}{cccccccccc}
		\toprule
		L   &     & 7.995 & 3.021 & 3.876 & 4.505 & 4.980 & 5.441 & 6.906 & 7.955 \\
		\midrule
		UCL &     & 8.7232 & 0.8096 & 1.2368 & 1.6515 & 2.0489 & 2.4666 & 4.5007 & 7.1431 \\
		\midrule
		$\lambda$ &     & 1   & 0.05 & 0.10 & 0.15 & 0.20 & 0.25 & 0.50 & 0.80 \\
		\midrule
		p   &     & Shewhart & EWMA & EWMA & EWMA & EWMA & EWMA & EWMA & EWMA \\
		\midrule
		0.40 & ARL & 499.64 & 500.24 & 500.19 & 500.13 & 500.25 & 500.44 & 502.34 & 494.97 \\
		& SDRL & 494.61 & 500.87 & 492.66 & 501.55 & 497.61 & 495.17 & 499.54 & 491.4 \\
		0.38 & ARL & 372.28 & \tikzhl{347.26} & 356.37 & 362.65 & 363.91 & 367.29 & 372.86 & 368.21 \\
		& SDRL & 376.29 & 342.59 & 349.72 & 360.15 & 360.72 & 363.64 & 374.97 & 371.91 \\
		0.35 & ARL & 240.71 & \tikzhl{210.20} & 217.09 & 224.35 & 227.43 & 231.69 & 238.16 & 238.32 \\
		& SDRL & 241.96 & 206.32 & 217.20 & 226.23 & 229.31 & 231.30 & 241.23 & 239.15 \\
		0.33 & ARL & 184.86 & \tikzhl{156.36} & 162.89 & 169.36 & 172.04 & 175.01 & 179.96 & 183.26 \\
		& SDRL & 186.15 & 150.71 & 162.44 & 168.63 & 172.04 & 175.59 & 181.53 & 184.16 \\ \hline
	\end{tabular}%
	\label{tab:7}
\end{table}%
\begin{table}[H]
	\centering
	\label{tab:8}
	\scriptsize
	\caption{Performance of ZINB EWMA and Shewhart charts for n=1, k=2, and $\theta$=0.80}
	\begin{tabular}{cccccccccc}
		\toprule
		L   &     & 6.647 & 2.863 & 3.575 & 4.067 & 4.457 & 4.786 & 5.960 & 6.445 \\
		\midrule
		UCL &     & 11.9972 & 1.3860 & 2.0062 & 2.5856 & 3.1473 & 3.7016 & 6.5000 & 9.6230 \\
		\midrule
		$\lambda$ &     & 1   & 0.05 & 0.10 & 0.15 & 0.20 & 0.25 & 0.50 & 0.80 \\
		\midrule
		p   &     & Shewhart & EWMA & EWMA & EWMA & EWMA & EWMA & EWMA & EWMA \\
		\midrule
		0.40 & ARL & 397.04 & 500.61 & 500.11 & 500.11 & 500.32 & 500.22 & 498.41 & 501.71 \\
		& SDRL & 402.61 & 500.11 & 498.45 & 505.46 & 511.13 & 509.21 & 499.68 & 508.79 \\
		0.38 & ARL & 275.51 & \tikzhl{315.78} & 326.24 & 330.37 & 330.98 & 333.84 & 338.47 & 346.38 \\
		& SDRL & 279.19 & 313.96 & 326.11 & 333.23 & 336.51 & 339.45 & 343.35 & 351.98 \\
		0.35 & ARL & 168.92 & \tikzhl{174.39} & 182.03 & 184.89 & 189.75 & 191.32 & 196.34 & 203.08 \\
		& SDRL & 166.81 & 169.03 & 178.29 & 180.67 & 186.46 & 187.61 & 196.29 & 204.58 \\
		0.33 & ARL & 122.45 & \tikzhl{123.41} & 128.62 & 131.51 & 133.58 & 134.66 & 139.17 & 144.88 \\
		& SDRL & 121.59 & 116.07 & 124.42 & 127.77 & 130.04 & 131.52 & 136.74 & 142.25 \\ \hline
	\end{tabular}%
	\label{tab:8}
\end{table}%
\begin{table}[H]
	\centering
	\scriptsize
	\label{tab:9}
	\caption{Performance of ZINB EWMA and Shewhart charts for n=1, k=5, and $\theta$=0.80}
	\begin{tabular}{cccccccccc}
		\toprule
		L   &     & 5.183 & 2.725 & 3.281 & 3.635 & 3.902 & 4.117 & 4.791 & 5.247 \\
		\midrule
		UCL &     & 20.007 & 3.0580 & 4.1877 & 5.1958 & 6.1443 & 7.0563 & 11.3768 & 16.7975 \\
		\midrule
		$\lambda$ &     & 1   & 0.05 & 0.10 & 0.15 & 0.20 & 0.25 & 0.50 & 0.80 \\
		\midrule
		p   &     & Shewhart & EWMA & EWMA & EWMA & EWMA & EWMA & EWMA & EWMA \\
		\midrule
		0.40 & ARL & 535.14 & 500.83 & 500.34 & 500.06 & 500.94 & 500.06 & 500.69 & 483.99 \\
		& SDRL & 526.63 & 497.33 & 497.26 & 497.27 & 495.76 & 495.30 & 510.74 & 476.21 \\
		0.38 & ARL & 319.13 & \tikzhl{280.43} & 289.95 & 291.71 & 291.29 & 295.22 & 294.36 & 287.60 \\
		& SDRL & 320.43 & 276.91 & 287.44 & 292.81 & 292.72 & 292.48 & 294.36 & 288.58 \\
		0.35 & ARL & 153.97 & 140.94 & 142.75 & 144.28 & 145.12 & 145.72 & 142.82 & \tikzhl{140.49} \\
		& SDRL & 153.55 & 135.42 & 139.19 & 142.02 & 143.53 & 143.72 & 141.02 & 141.05 \\
		0.33 & ARL & 101.03 & 95.05 & 96.19 & 96.35 & 96.37 & 95.60 & 92.81 & \tikzhl{92.19} \\
		& SDRL & 99.62 & 88.14 & 91.93 & 94.06 & 94.55 & 94.35 & 89.79 & 90.64 \\ \hline
	\end{tabular}%
	\label{tab:9}
\end{table}%

\begin{table}[H]
	\centering
	\label{tab:13}
	\scriptsize
	\caption{Performance of ZINB EWMA and Shewhart charts for n=1, k=1, and $\theta$=0.75}
	\begin{tabular}{cccccccccc}
		\toprule
		L   &     & 7.395 & 2.944 & 3.749 & 4.316 & 4.771 & 5.151 & 6.407 & 7.17 \\
		\midrule
		UCL &     & 8.9969 & 0.9246 & 1.3777 & 1.8078 & 2.2292 & 2.6449 & 4.6878 & 7.2006 \\
		\midrule
		$\lambda$ &     & 1   & 0.05 & 0.10 & 0.15 & 0.20 & 0.25 & 0.50 & 0.80 \\
		\midrule
		p   &     & Shewhart & EWMA & EWMA & EWMA & EWMA & EWMA & EWMA & EWMA \\
		\midrule
		0.40 & ARL & 405.29 & 500.21 & 500.61 & 500.31 & 500.11 & 500.52 & 500.51 & 496.17 \\
		& SDRL & 402.91 & 498.47 & 501.05 & 501.17 & 507.55 & 507.87 & 503.53 & 499.86 \\
		0.38 & ARL & 297.34 & \tikzhl{333.91} & 345.67 & 353.29 & 355.94 & 359.15 & 365.27 & 361.65 \\
		& SDRL & 298.98 & 329.52 & 349.67 & 358.57 & 361.72 & 364.54 & 368.07 & 362.72 \\
		0.35 & ARL & 192.58 & \tikzhl{192.51} & 203.22 & 209.61 & 215.73 & 219.63 & 228.44 & 227.15 \\
		& SDRL & 194.78 & 187.30 & 199.36 & 207.17 & 215.34 & 219.83 & 229.59 & 230.98 \\
		0.33 & ARL & 147.42 & \tikzhl{141.09} & 150.71 & 155.25 & 159.56 & 163.34 & 169.97 & 170.81 \\
		& SDRL & 147.15 & 135.18 & 146.28 & 152.07 & 157.34 & 161.46 & 168.83 & 170.56 \\ \hline
	\end{tabular}%
	\label{tab:13}
\end{table}%
\begin{table}[H]
	\centering
	\scriptsize
	\label{tab:14}
	\caption{Performance of ZINB EWMA and Shewhart charts for n=1, k=2, and $\theta$=0.75}
	\begin{tabular}{cccccccccc}
		\toprule
		L   &     & 6.295 & 2.797 & 3.465 & 3.908 & 4.254 & 4.548 & 5.525 & 6.261 \\
		\midrule
		UCL &     & 12.6315 & 1.5953 & 2.2503 & 2.8503 & 3.4264 & 3.9945 & 6.7707 & 10.3988 \\
		\midrule
		$\lambda$ &     & 1   & 0.05 & 0.10 & 0.15 & 0.20 & 0.25 & 0.50 & 0.80 \\
		\midrule
		p   &     & Shewhart & EWMA & EWMA & EWMA & EWMA & EWMA & EWMA & EWMA \\
		\midrule
		0.40 & ARL & 497.27 & 500.79 & 500.92 & 500.09 & 500.13 & 500.54 & 500.84 & 479.06 \\%
		& SDRL & 498.81 & 493.09 & 494.11 & 498.38 & 496.58 & 500.17 & 503.21 & 479.09 \\%
		0.38 & ARL & 336.73 & \tikzhl{300.97} & 314.31 & 317.86 & 319.81 & 323.96 & 335.61 & 325.19 \\
		& SDRL & 342.13 & 298.76 & 309.56 & 311.91 & 317.61 & 322.94 & 337.15 & 329.52 \\
		0.35 & ARL & 195.51 & \tikzhl{159.35} & 169.59 & 173.49 & 176.41 & 179.63 & 188.44 & 188.44 \\
		& SDRL & 196.63 & 153.24 & 166.78 & 170.87 & 172.86 & 175.34 & 187.55 & 188.69 \\
		0.33 & ARL & 138.01 & \tikzhl{110.07} & 116.43 & 119.81 & 121.36 & 123.46 & 131.35 & 132.39 \\
		& SDRL & 135.21 & 102.06 & 111.01 & 115.57 & 117.68 & 119.27 & 128.01 & 129.56 \\ \hline
	\end{tabular}%
	\label{tab:14}
\end{table}%
\begin{table}[H]
	\centering
	\scriptsize
	\label{tab:15}
	\caption{Performance of ZINB EWMA and Shewhart charts for n=1, k=5, and $\theta$=0.75}
	\begin{tabular}{cccccccccc}
		\toprule
		L   &     & 4.898 & 2.663 & 3.175 & 3.496 & 3.724 & 3.906 & 4.493 & 4.824 \\
		\midrule
		UCL &     & 20.9925 & 3.5393 & 4.718 & 5.7604 & 6.72 & 7.6372 & 11.9998 & 17.2485 \\
		\midrule
		$\lambda$ &     & 1   & 0.05 & 0.10 & 0.15 & 0.20 & 0.25 & 0.50 & 0.80 \\
		\midrule
		p   &     & Shewhart & EWMA & EWMA & EWMA & EWMA & EWMA & EWMA & EWMA \\
		\midrule
		0.40 & ARL & 423.36 & 500.18 & 500.37 & 500.75 & 500.92 & 500.06 & 499.45 & 500.16 \\
		& SDRL & 416.96 & 504.01 & 498.21 & 499.45    & 503.98 & 505.36 & 501.96 & 498.25 \\
		0.38 & ARL & 251.43 & \tikzhl{261.43} & 271.52 & 276.67 & 276.45 & 277.61 & 283.84 & 290.72 \\
		& SDRL & 248.12 & 256.87 & 268.92 & 279.42 & 277.22 & 279.81 & 283.54 & 290.93 \\
		0.35 & ARL & 123.51 & \tikzhl{126.19} & 129.86 & 131.58 & 132.42 & 132.61 & 133.07 & 139.23 \\
		& SDRL & 121.81 & 117.94 & 126.44 & 129.27 & 130.96 & 130.60 & 131.48 & 137.81 \\
		0.33 & ARL & 80.47 & \tikzhl{84.08} & 85.03 & 85.88 & 85.56 & 85.56 & 84.49 & 88.72 \\
		& SDRL & 79.05 & 76.81 & 79.76 & 82.66 & 83.06 & 83.03 & 82.80 & 86.74 \\ \hline
	\end{tabular}%
	\label{tab:15}
\end{table}%
	\begin{table}[H]
	\centering
	\scriptsize
	\label{tab:1n}
	\caption{Performance of ZINB EWMA and Shewhart charts for n=10, k=1, and $\theta$=0.85}
	\begin{tabular}{rccccccccc}
		\toprule
		\multicolumn{1}{l}{L} &     & 4.891 & 2.592 & 3.083 & 3.391 & 3.631 & 3.830 & 4.502 & 4.887 \\
		\midrule
		\multicolumn{1}{l}{UCL} &     & 1.6504 & 0.3459 & 0.4311 & 0.5064 & 0.5777 & 0.6468 & 0.9825 & 1.3879 \\
		\midrule
		\multicolumn{1}{l}{$\lambda$} &     & 1   & 0.05 & 0.10 & 0.15 & 0.20 & 0.25 & 0.50 & 0.80 \\
		\midrule
		\multicolumn{1}{c}{p} &     & Shewhart & EWMA & EWMA & EWMA & EWMA & EWMA & EWMA & EWMA \\
		\midrule
		\multicolumn{1}{c}{0.40} & ARL & 449.98 & 500.24 & 500.89 & 500.77 & 501.21 & 500.24 & 500.99 & 499.47 \\
		& SDRL & 455.44 & 497.67 & 497.11 & 498.31 & 497.01 & 494.39 & 498.12 & 501.31 \\
		\multicolumn{1}{c}{0.38} & ARL & 290.22 & \tikzhl{221.29} & 244.25 & 255.38 & 266.82 & 275.52 & 301.39 & 314.87 \\
		& SDRL & 286.91 & 213.04 & 239.46 & 253.43 & 264.79 & 269.99 & 297.01 & 310.88 \\
		\multicolumn{1}{c}{0.35} & ARL & 151.83 & \tikzhl{87.66} & 99.79 & 108.65 & 114.92 & 121.39 & 143.87 & 158.99 \\
		& SDRL & 150.21 & 79.61 & 95.78 & 106.27 & 114.51 & 119.91 & 143.19 & 157.04 \\
		\multicolumn{1}{c}{0.33} & ARL & 103.27 & \tikzhl{54.82} & 61.06 & 66.85 & 71.57 & 75.91 & 93.57 & 105.91 \\
		& SDRL & 101.81 & 46.29 & 56.04 & 63.59 & 69.10 & 74.03 & 91.15 & 104.48 \\
		\bottomrule
	\end{tabular}%
	\label{tab:1n}%
\end{table}%

\begin{table}[H]
	\centering
	\scriptsize
	\label{tab:2n}
	\caption{Performance of ZINB EWMA and Shewhart charts for n=10, k=1, and $\theta$=0.80}
	\begin{tabular}{rccccccccc}
		\toprule
		\multicolumn{1}{l}{L} &     & 4.625 & 2.564 & 3.014 & 3.297 & 3.514 & 3.687 & 4.275 & 4.621 \\
		\midrule
		\multicolumn{1}{l}{UCL} &     & 1.8408 & 0.4367 & 0.5303 & 0.6127 & 0.6902 & 0.7642 & 1.1223 & 1.557 \\
		\midrule
		\multicolumn{1}{l}{$\lambda$} &     & 1   & 0.05 & 0.10 & 0.15 & 0.20 & 0.25 & 0.50 & 0.80 \\
		\midrule
		\multicolumn{1}{c}{p} &     & Shewhart & EWMA & EWMA & EWMA & EWMA & EWMA & EWMA & EWMA \\
		\midrule
		\multicolumn{1}{c}{0.40} & ARL & 469.15 & 500.66 & 500.02 & 500.41 & 500.66 & 501.01 & 500.43 & 499.08 \\
		& SDRL & 468.62 & 490.84 & 497.12 & 498.61 & 498.58 & 506.06 & 502.84 & 497.78 \\
		\multicolumn{1}{c}{0.38} & ARL & 288.78 & \tikzhl{201.63} & 220.11 & 234.59 & 247.47 & 255.83 & 280.73 & 297.91 \\
		& SDRL & 290.91 & 191.35 & 216.23 & 230.71 & 242.66 & 251.49 & 278.41 & 296.24 \\
		\multicolumn{1}{c}{0.35} & ARL & 146.13 & \tikzhl{74.09} & 82.69 & 91.24 & 98.84 & 105.24 & 127.81 & 145.14 \\
		& SDRL & 147.26 & 65.53 & 77.88 & 88.41 & 97.29 & 104.46 & 128.89 & 145.94 \\
		\multicolumn{1}{c}{0.33} & ARL & 94.34 & \tikzhl{45.39} & 49.89 & 54.29 & 58.99 & 62.51 & 78.23 & 92.21 \\
		& SDRL & 94.71 & 36.92 & 44.88 & 50.46 & 56.01 & 60.01 & 77.92 & 93.11 \\
		\bottomrule
	\end{tabular}%
	\label{tab:2n}%
\end{table}%
\begin{table}[H]
	\centering
	\scriptsize
	\label{tab:3n}
	\caption{Performance of ZINB EWMA and Shewhart charts for n=10, k=1, and $\theta$=0.75}
	\begin{tabular}{rccccccccc}
		\toprule
		\multicolumn{1}{l}{L} &     & 4.449 & 2.536 & 2.968 & 3.220 & 3.416 & 3.572 & 4.119 & 4.442 \\
		\midrule
		\multicolumn{1}{l}{UCL} &     & 2.0153 & 0.5247 & 0.6260 & 0.7130 & 0.7948 & 0.8727 & 1.2518 & 1.7122 \\
		\midrule
		\multicolumn{1}{l}{$\lambda$} &     & 1   & 0.05 & 0.10 & 0.15 & 0.20 & 0.25 & 0.50 & 0.80 \\
		\midrule
		\multicolumn{1}{c}{p} &     & Shewhart & EWMA & EWMA & EWMA & EWMA & EWMA & EWMA & EWMA \\
		\midrule
		\multicolumn{1}{c}{0.40} & ARL & 525.03 & 500.22 & 500.42 & 500.06 & 500.57 & 500.19 & 500.04 & 500.00 \\
		& SDRL & 523.16 & 493.31 & 496.74 & 504.8 & 507.52 & 504.03 & 495.89 & 500.65 \\
		\multicolumn{1}{c}{0.38} & ARL & 307.31 & \tikzhl{183.14} & 206.58 & 218.87 & 231.62 & 237.55 & 269.68 & 284.92 \\
		& SDRL & 310.62 & 175.99 & 201.53 & 214.31 & 226.81 & 234.83 & 271.32 & 286.91 \\
		\multicolumn{1}{c}{0.35} & ARL & 147.82 & \tikzhl{63.75} & 73.02 & 79.92 & 86.71 & 92.51 & 116.99 & 134.43 \\
		& SDRL & 147.07 & 54.49 & 67.85 & 76.58 & 83.91 & 90.43 & 116.85 & 135.14 \\
		\multicolumn{1}{c}{0.33} & ARL & 94.71 & \tikzhl{38.79} & 42.41 & 46.27 & 50.01 & 53.67 & 70.38 & 84.92 \\
		& SDRL & 94.59 & 30.69 & 36.65 & 42.09 & 47.05 & 51.15 & 70.05 & 84.23 \\
		\bottomrule
	\end{tabular}%
	\label{tab:3n}%
\end{table}%

Tables \ref{tab:1n}, \ref{tab:2n}, and \ref{tab:3n} show the ARL and SDRL of the ZINB-EWMA and ZINB-Shewhart charts for mean monitoring. For example, Table \ref{tab:1n} shows the results assuming IC process with parameters (n, k, $p_0$, $\theta$)=(10, 1, 0.40, 0.85), and $\lambda$=0.05. For the EWMA chart, L=2.592 produces an $ARL_0$=500.24 with $SDRL_0$=497.67. For the ZINB-EWMA, when (n, k, $p_0$, $\theta$)=(10, 1, 0.40, 0.85) with $\lambda$=0.15 and L=3.391, gives the value of $ARL_0$= 500.77 and $SDRL_0$=498.31. Similarly, for (n, k, $p_0$, $\theta$)=(1, 1, 0.40, 0.85) with $\lambda$=1 and L=4.891, the ZINB-Shewhart produces $ARL_0$=449.98 with $SDRL_0$=455.44. For the OOC process, when (n, k, $p_0$, $\theta$)=(10, 1, 0.40, 0.85) and $p_1$=0.38, the $ARL_1$ and $SDRL_1$ for the ZINB-EWMA chart with $\lambda$=0.05 and L=2.592 are 221.29 and 213.04, respectively. Assuming (n, k, $p_0$, $\theta$)=(10, 1, 0.40, 0.85) and $p_1$=0.38, the $ARL_1$ and $SDRL_1$ for the ZINB-EWMA chart with $\lambda$=0.15 and L=3.391 are 255.38 and 253.43, respectively. Similarly, we calculate the $ARL_1$ and $SDRL_1$ of the ZINB-EWMA for the different values of $\lambda$ like (0.10,0.20,0.25,0.50,0.80). Assuming (n, k, $p_0$, $\theta$)=(10, 1, 0.40, 0.85) and $p_1$=0.38, $ARL_1$ and $SDRL_1$ for the ZINB-Shewhart chart with $\lambda$=1 and L=4.891 are 290.22 and 286.91, respectively. However, it is worth mentioning that the IC ARL is not achieved for the Shewhart chart. For $p_1$=0.35 and (n, k, $p_0$, $\theta$)=(10, 1, 0.40, 0.85) the $ARL_1$ and $SDRL_1$ of the ZINB-Shewhart with $\lambda$=1 and L=4.891 are 151.83 and 150.21, respectively. The results reported in Tables \ref{tab:2n} and \ref{tab:3n} for $\theta$ =0.80, 0.75 with k=1 and n=10 can be interpreted similarly.




One can notice that the ZINB-EWMA performs well as compared to the ZINB-Shehart because the ZINB-EWMA achieved the desire IC ARL in all cases of $\lambda$. Also, it is observed that the $L$ values decreased by decreasing the value of $\theta$ and vice versa.
	\renewcommand{\arraystretch}{0.95}
	\renewcommand\tabcolsep{1.0 pt}
\begin{table}[htbp]
	\centering
	\small
	\caption{ZINB-EWMA ($\lambda=0.10$) and ZINB-Shewhart ($\lambda=1.0$) for $n=10$ assuming shifts in all parameters and $p_0=0.4,\theta_0=0.85,k_0=1$}
	\begin{tabular}{|ccc|cc|cc|cc|}
		\toprule
		& &&\multicolumn{2}{c|}{Shewhart, L=4.891 ($ARL_0=500$)} & \multicolumn{2}{c|}{EWMA L=3.083 ($ARL_0=500$)} & \multicolumn{2}{c|}{EWMA L=3.012 ($ARL_0=449$)} \\
		\hline
		$p_1$ & $\theta_1$ & $k_1$   & ARL & SDLR & ARL & SDLR & ARL & SDLR \\
		\hline
		{\bf0.4} & {\bf0.85} & {\bf1}   & {\bf449.98} & 455.44 & {\bf500.89} & 497.11 & {\bf449.8} & 445.58 \\
		&     & 2   & 860.23 & 853.12 & 645.99 & 644.17 & 569.59 & 563.99 \\
		&     & 3   & 1421.47 & 1389.54 & 729.08 & 718.08 & 638.13 & 619.74 \\
		&     & 5   & 2589.76 & 2244.79 & 826.62 & 827.28 & 718.52 & 715.67 \\
		0.4 & 0.8 & 1   & 681.47 & 677.42 & 560.77 & 562.23 & 498.55 & 495.86 \\
		&     & 2   & 1417.45 & 1396.78 & 726.05 & 712.18 & 638.07 & 630.25 \\
		&     & 3   & 2323.69 & 2097.86 & 805.82 & 803.18 & 700.55 & 697.35 \\
		&     & 5   & 3377.83 & 2612.13 & 925.55 & 914.91 & 794.12 & 792.22 \\
		0.4 & 0.75 & 1   & 751.18 & 748.67 & 614.19 & 604.74 & 539.45 & 534.94 \\
		&     & 2   & 1862.39 & 1752.51 & 789.35 & 797.89 & 690.02 & 690.41 \\
		&     & 3   & 3040.55 & 2485.81 & 910.92 & 902.411 & 782.82 & 772.28 \\
		&     & 5   & 3919.08 & 2767.21 & 1040.33 & 1012.83 & 888.79 & 861.55 \\
		0.38 & 0.85 & 1   & 299.22 & 286.91 & 244.25 & 239.46 & 223.16 & 218.13 \\
		&     & 2   & 481.71 & 480.88 & 252.32 & 249.02 & 229.29 & 226.76 \\
		&     & 3   & 721.13 & 711.92 & 256.31 & 251.91 & 231.56 & 225.58 \\
		&     & 5   & 1226.63 & 1227.15 & 259.03 & 252.38 & 234.05 & 228.97 \\
		0.38 & 0.8 & 1   & 407.79 & 409.72 & 241.37 & 234.28 & 219.92 & 216.19 \\
		&     & 2   & 724.55 & 740.84 & 247.58 & 236.64 & 223.64 & 215.37 \\
		&     & 3   & 1126.52 & 1131.23 & 245.51 & 238.12 & 218.68 & 211.23 \\
		&     & 5   & 1760.13 & 1689.67 & 242.35 & 230.71 & 218.74 & 206.18 \\
		0.38 & 0.75 & 1   & 428.35 & 433.43 & 240.36 & 233.54 & 218.89 & 214.02 \\
		&     & 2   & 907.59 & 928.04 & 241.64 & 232.91 & 215.75 & 207.79 \\
		&     & 3   & 1565.89 & 1525.63 & 233.83 & 223.79 & 208.71 & 200.11 \\
		&     & 5   & 2410.41 & 2164.52 & 226.97 & 217.76 & 204.12 & 195.61 \\
		{0.35} & {0.85} & 1   & 151.83 & 150.21 & 99.79 & 95.78 & 92.31 & 88.02 \\
		&     & 2   & 210.66 & 210.23 & 86.86 & 79.85 & 80.81 & 74.47 \\
		&     & 3   & 278.21 & 282.48 & 80.33 & 72.89 & 74.68 & 67.35 \\
		&     & 5   & 396.11 & 399.07 & 73.26 & 66.88 & 67.99 & 61.81 \\
		{0.35} & {0.8} & 1   & 196.74 & 195.54 & 88.93 & 83.53 & 82.51 & 77.66 \\
		&     & 2   & 278.87 & 274.55 & 75.59 & 69.14 & 70.07 & 63.56 \\
		&     & 3   & 379.46 & 285.42 & 68.62 & 62.36 & 63.58 & 57.44 \\
		&     & 5   & 496.49 & 498.73 & 60.23 & 53.59 & 56.27 & 49.84 \\
		{0.35} & {0.75} & 1   & 198.63 & 199.41 & 81.59 & 76.28 & 76.18 & 70.94 \\
		&     & 2   & 319.77 & 315.05 & 66.08 & 59.24 & 61.62 & 55.36 \\
		&     & 3   & 466.15 & 458.31 & 58.63 & 51.06 & 54.71 & 47.26 \\
		&     & 5   & 610.98 & 606.36 & 52.06 & 44.42 & 48.63 & 41.38 \\
		{0.33} & {0.85} & 1   & 103.27 & 101.81 & 61.06 & 56.03 & 57.55 & 52.55 \\
		&     & 2   & 127.58 & 126.78 & 50.11 & 44.01 & 47.47 & 41.73 \\
		&     & 3   & 151.26 & 151.08 & 45.48 & 39.87 & 42.72 & 37.25 \\
		&     & 5   & 200.32 & 200.72 & 40.11 & 34.12 & 37.91 & 32.33 \\
		{0.33} & {0.8} & 1   & 124.32 & 123.97 & 52.67 & 47.41 & 49.8 & 44.83 \\
		&     & 2   & 156.76 & 155.02 & 41.95 & 35.89 & 39.71 & 34.11 \\
		&     & 3   & 193.45 & 194.37 & 37.56 & 31.42 & 35.45 & 29.51 \\
		&     & 5   & 229.85 & 230.81 & 32.81 & 26.57 & 31.01 & 25.18 \\
		{0.33} & {0.75} & 1   & 122.97 & 124.23 & 46.87 & 40.81 & 44.06 & 38.29 \\
		&     & 2   & 169.53 & 168.57 & 36.05 & 29.97 & 34.03 & 28.09 \\
		&     & 3   & 225.84 & 225.74 & 31.39 & 25.17 & 29.81 & 23.87 \\
		&     & 5   & 260.15 & 260.25 & 27.42 & 21.43 & 26.14 & 20.29 \\
		\bottomrule
	\end{tabular}%
	\label{tab:4}%
\end{table}%
\begin{table}[htbp]
	\centering
	\small
	\caption{ZINB-EWMA ($\lambda=0.10$) and ZINB-Shewhart ($\lambda=1.0$) for $n=10$ assuming shifts in all parameters and $p_0=0.4,\theta_0=0.85,k_0=1$}
	\begin{tabular}{|ccc|cc|cc|cc|}
		\toprule
		& &&\multicolumn{2}{c|}{Shewhart, L=4.891 ($ARL_0=500$)} & \multicolumn{2}{c|}{EWMA L=3.083 ($ARL_0=500$)} & \multicolumn{2}{c|}{EWMA L=3.012 ($ARL_0=449$)} \\
		\hline
		$p_1$ & $\theta_1$ & $k_1$   & ARL & SDLR & ARL & SDLR & ARL & SDLR \\
		\hline
		{\bf0.4} & {\bf0.85} & {\bf1}   & {\bf449.98} & 455.44 & {\bf500.89} & 497.11 & {\bf449.8} & 445.58 \\
		&     & 2   &38.13  &37.61  &16.14  &11.76  &15.60  &11.39  \\
		&     & 3   &11.56  &10.94  &6.97  &4.43  &6.81  &4.34  \\
		&     & 5   &3.80  &3.22  &3.55  &2.14  &3.49  &2.10  \\
		0.4 & 0.8 & 1   &222.97  &223.36  &83.07  &77.35  &77.15  & 72.10 \\
		&     & 2   &19.93  &19.48  &8.60  &5.27  &8.36  &5.12  \\
		&     & 3  &6.55  &5.99  &4.52  &2.55  &4.43  &2.49  \\
		&     & 5  &2.52  &1.95  &2.56  &1.41  &2.53  &1.39  \\
		0.4 & 0.75 & 1   &124.85  &125.06  &30.93  &24.91  &29.41  &23.70  \\
		&     & 2   &12.06  &11.66  &5.83  &3.18  &5.71  &3.13  \\
		&     & 3   &4.28  &3.68  &3.39  &1.75  & 3.34 &1.72  \\
		&     & 5   &1.9  &1.3  &2.04  &1.03  &2.03  &1.01  \\
		0.38 & 0.85 & 1   &290.22  &286.90  &244.25  &239.46  &223.16  &218.14  \\
		&     & 2   &28.04  &27.49  &13.09  &9.34  &12.70  &9.05  \\
		&     & 3   &9.05  &8.53  &6.11  &3.84  &5.99  &3.78  \\
		&     & 5  &3.30  &2.75  &3.23  &1.97  &3.19  &1.93  \\
		0.38 & 0.8 & 1   &146.03  &146.43  &52.41  &46.11  &49.32  &43.37  \\
		&     & 2   &15.05  &14.27  &7.36  &4.43  &7.17  &4.33  \\
		&     & 3   &5.35  &4.78  &4.08  & 2.31 &4.01  &2.26  \\
		&     & 5   &2.27  &1.71  &2.37  &1.31  &2.36  &1.29  \\
		0.38 & 0.75 & 1   &84.22  &82.99  &22.49  &16.98  &21.52  &16.15  \\
		&     & 2   &9.33  &8.76  &5.14  &2.78  &5.03  &2.73  \\
		&     & 3  &3.60  &3.06  &3.08  &1.60  &3.04  &1.58  \\
		&     & 5  &1.75  &1.15  &1.91  &0.96  &1.90  &0.95  \\
		{0.35} & {0.85} & 1   &151.83  &150.21  &99.79  &95.79  &92.31  &88.02  \\
		&     & 2   &18.16  &17.78  &9.92  &6.80  &9.64  &6.59  \\
		&     & 3   &6.61  &6.06  &5.08  &3.18  &4.98  &3.11  \\
		&     & 5   &2.80  &2.25  &2.85  &1.75  &2.82  &1.70  \\
		{0.35} & {0.8} & 1   &80.06  &80.28  &29.68  &24.28  &28.36  &23.12  \\
		&     & 2   &10.14  &9.53  &5.96  &3.54  &5.85  &3.48  \\
		&     & 3   &4.11  &3.58  &3.51  &2.00  &3.45  &1.96  \\
		&     & 5   &1.98  &1.40  &2.13  &1.18  &2.11  &1.16  \\
		{0.35} & {0.75} & 1   &46.59  &46.44  &15.42  &10.72  &14.86  &10.31  \\
		&     & 2   &6.52  &5.94  &4.32  &2.32  &4.24  &2.28  \\
		&     & 3   &2.86  &2.30  &2.71  &1.41  &2.68  &1.38  \\
		&     & 5  &1.58  &0.97  &1.74  &0.87  &1.74  &0.86  \\
		{0.33} & {0.85} & 1   &103.28  &101.80  &61.06  &56.04  &57.56  &52.56  \\
		&     & 2   &13.84  &13.36  &8.38  &5.60  &8.15  &5.43  \\
		&     & 3   &5.45  &4.85  &4.54  &2.84  &4.46  &2.78  \\
		&     & 5   & 2.52 &1.98  &2.64  &1.62  &2.61  &1.59  \\
		{0.33} & {0.8} & 1   &54.39  &54.76  &21.72  &16.88  &20.86  &16.08  \\
		&     & 2   &7.98  &7.36  &5.26  &3.08  &5.15  &3.02  \\
		&     & 3  &3.49  &2.94  &3.20  &1.82  &3.16  &1.77  \\
		&     & 5   &1.84  &1.26  &1.99  &1.11  &1.99  &1.09  \\
		{0.33} & {0.75} & 1   &33.05  &32.89  &12.26  &8.06  &11.88  &7.82  \\
		&     & 2  &5.23  &4.61  &3.85  &2.08  &3.79  &2.04  \\
		&     & 3   &2.51  &1.94  &2.50  &1.29  &2.48  &1.27  \\
		&     & 5   &1.48  &0.85  &1.65  &0.81  &1.64  &0.80  \\
		\bottomrule
	\end{tabular}%
	\label{tab:4n}%
\end{table}%

In the previous tables, we studied the performance of the charts assuming shifts in the parameter $p$. In practice, a shift can appear in any parameter of the ZINB distribution. To show the performance for this case, Table \ref{tab:4n} show the ARL and SDRL values of the ZINB-EWMA and the ZINB-Shewhart for $n=10$ with $ARL_0=500$ (fourth and sixth columns) and $ARL_0=449$ (eighth column to match with the $ARL_0$ achieved by the Shewhart chart in the fourth column). Table \ref{tab:4n} tabulates the results for mean monitoring using Shewhart and EWMA charts. Again, it is noticed that the $ARL_0$ is not achieved for the Shewhart chart. The $ARL_1$ for EWMA and Shewhart charts decreased as $p_0$ shifted to $p_1$. However, the EWMA chart outperform the Shewhart chart. In particular, the IC ARL of the ZINB-EWMA chart with parameters (n, k, $p_0$, $\theta$)=(10, 1, 0.40, 0.85), $\lambda$=0.10, and L=3.083 is $ARL_0$=500.89 with $SDRL_0$=497.11. For the Shewhart chart assuming (n, k, $p_0$, $\theta$)=(10, 1, 0.40, 0.85) with $\lambda$=1 and L=4.891, the IC process will give the $ARL_0$=449.98 with $SDRL_0$=455.44. For the OOC specifications (n, k, $p_1$, $\theta$)=(10, 1, 0.40, 0.80) and $p_1$=0.40, the $ARL_1$ and $SDRL_1$ for the ZINB-EWMA chart with $\lambda$=0.1 and L=3.083 are 83.07 and 77.35, respectively. However, for (n, k, $p_1$, $\theta$)=(10, 1, 0.40, 0.75) and $p_1$=0.38, the $ARL_1$ for the ZINB-EWMA chart with $\lambda$=0.10 and L=3.083 are 30.93 and 244.25, respectively. For the ZINB-Shewhart chart with (n, k, $p_1$, $\theta$)=(10, 2, 0.40, 0.85), $\lambda$=1, and L=4.891, $ARL_1=38.13$ and $SDRL_1=37.61$. Similarly, when $p_1$=0.38 and (n, k, $p_0$, $\theta$)=(10, 1, 0.40, 0.85), the $ARL_1$ and $SDRL_1$ of the ZINB-Shewhart with $\lambda$=1 and L=4.891 are 290.22 and 286.90, respectively. Generally, it is noticed that the ARL values of both charts decrease as the values of $k$ increased. The ZINB-EWMA performs well as compared to the ZINB-Shewhart, because the ZINB-EWMA give small value of $ARL_1$ in all cases of $p$ and $k$.

The last two columns of Table \ref{tab:4n} list the ARL and SDRL values of the ZINB-EWMA for $n=10$ assuming $ARL_0=449$ as this IC ARL is achieved in the 3rd column of the same table. It is again evident that the EWMA chart outperforms the Shewhart chart. 

In the previous tables, we studied the performance of the charts assuming shifts in the parameter $p$. In practice, a shift can appear in any parameter of the ZINB distribution. To show the performance for this case, Table \ref{tab:4} show the ARL and SDRL values of the ZINB-EWMA and the ZINB-Shewhart for $n=10$ with $ARL_0=500$ (fouth and sixth columns) and $ARL_0=449$ (eighth column to match with the $ARL_0$ achieved by the Shewhart chart in the fourth column). Table \ref{tab:4} tabulates the results for mean monitoring using Shewhart and EWMA charts. Again, it is noticed that the $ARL_0$ is not achieved for the Shewhart chart. For $p_0=0.40$, the $ARL_1$ increases as the value of $k$ increased, especially for the Shewhart chart. However, the $ARL_1$ for EWMA chart decreased as $p_0$ shifted to $p_1$. Thus, the EWMA chart is more efficient than the Shewhart chart. In particular, the IC ARL of the ZINB-EWMA chart with parameters (n, k, $p_0$, $\theta$)=(10, 1, 0.40, 0.85), $\lambda$=0.10, and L=3.083 is $ARL_0$=500.89 with $SDRL_0$=497.11. For the Shewhart chart assuming (n, k, $p_0$, $\theta$)=(10, 1, 0.40, 0.85) with $\lambda$=1 and L=4.891, the IC process will give the $ARL_0$=449.98 with $SDRL_0$=455.44. For the OOC specifications (n, k, $p_1$, $\theta$)=(10, 1, 0.40, 0.80) and $p_1$=0.40, the $ARL_1$ and $SDRL_1$ for the ZINB-EWMA chart with $\lambda$=0.1 and L=3.083 are 560.77 and 562.23, respectively. However, for (n, k, $p_1$, $\theta$)=(10, 1, 0.40, 0.75) and $p_1$=0.38, the $ARL_1$ and $SDRL_1$ for the ZINB-EWMA chart with $\lambda$=0.10 and L=3.083 are 240.36 and 233.54, respectively. For the ZINB-Shewhart chart with (n, k, $p_1$, $\theta$)=(10, 2, 0.40, 0.85), $\lambda$=1, and L=4.891, $ARL_1=860.23$ and $SDRL_1=853.12$. Similarly, when $p_1$=0.38 and (n, k, $p_0$, $\theta$)=(10, 1, 0.40, 0.85), the $ARL_1$ and $SDRL_1$ of the ZINB-Shewhart with $\lambda$=1 and L=4.891 are 299.22 and 286.91, respectively. Generally, it is noticed that the ARL values of the ZINB-Shewhart increase as the values of $k$ increased. However, for the ZINB-EWMA shows a decreasing pattern as a shift is introduced in $p$. The ZINB-EWMA performs well as compared to the ZINB-Shewhart, because the ZINB-EWMA give small value of $ARL_1$ in all cases of $p$ and $k$.

The last two columns of Table \ref{tab:4} list the ARL and SDRL values of the ZINB-EWMA for $n=10$ assuming $ARL_0=449$ as this IC ARL is achieved in the 3rd column of the same table. It is again evident that the EWMA chart outperforms the Shewhart chart. 
\subsection{Comparison between EWMA and Shewhart charts}
In this section, a comparison of the ZINB-EWMA and ZINB-Shewhart is discussed. For the comparison of the EWMA and Shewhart, we set $ARL_0=500$ and 370 and noticed from the tables that the ZINB-Shewhart control charts give the $ARL_0$ value, which is not exactly 500 when the process is IC. Thus, $ARL_0$ is not achieved in the Shewhart chart and $ARL_1$ results, therefore, are not trustworthy. A very slight change in the value of $L$ gives a very large value of $ARL_0$ of the ZINB-Shewhart. For example, in Table \ref{tab:1}, the $ARL_0$ value of ZINB-Shewhart with $\lambda=1$ and L = 5.067 is 396.01. But when we use $L=5.949$ (Table \ref{tab:3}), it gives the $ARL_0$ 499.02 with SDRL 494.64. 

For the ZINB-EWMA, different values of $\lambda$ such as (0.05,0.10,0.15,0.20,0.25,0.50,.0.80) are used. The ZINB-EWMA gives the pre-fixed value of $ARL_0$ in all cases of $\lambda$ values. It also gives the smaller value of $ARL_1$ in all cases when k=1, 2, 3, 5, and $\theta$=0.85, 0.80, 0.75, except when all parameters have shifts and particularly $k$ is large. Hence, the ZINB-EWMA indicates better results as compared to the ZINB-Shewhart control chart. 

\begin{figure}[h]
	\centering
	\begin{subfigure}[b]{0.48\textwidth}
		\centering
		\includegraphics[width=\textwidth]{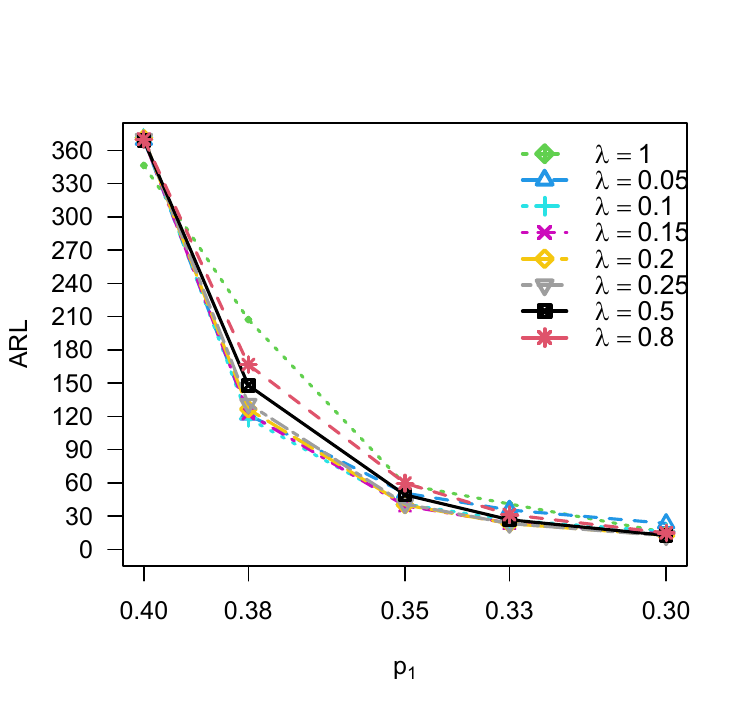}
		\caption{$k=1$}
		\label{3.2a}
	\end{subfigure}
	\hfill
	\begin{subfigure}[b]{0.48\textwidth}
		\centering
		\includegraphics[width=\textwidth]{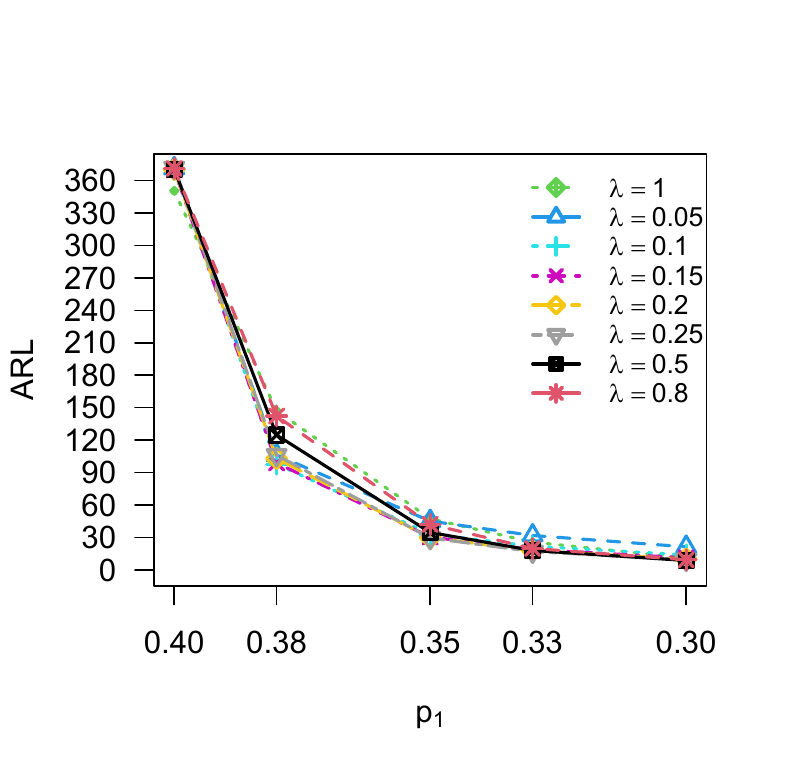}
		\caption{$k=2$}
		\label{3.2b}
	\end{subfigure}
	\hfill
	\begin{subfigure}[b]{0.48\textwidth}
		\centering
		\includegraphics[width=\textwidth]{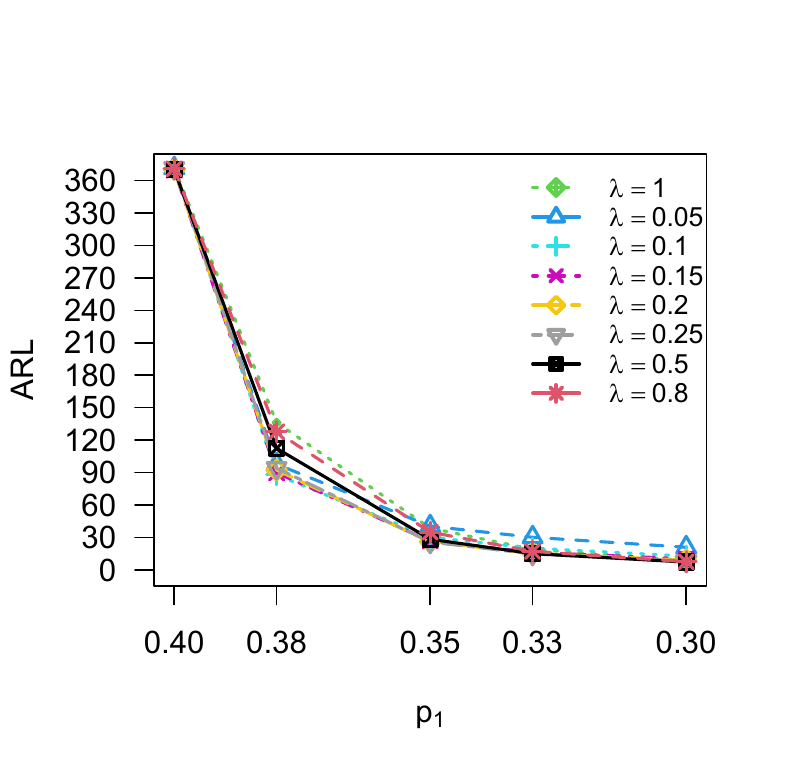}
		\caption{$k=3$}
		\label{3.2c}
	\end{subfigure}
	\hfill
	\begin{subfigure}[b]{0.48\textwidth}
		\centering
		\includegraphics[width=\textwidth]{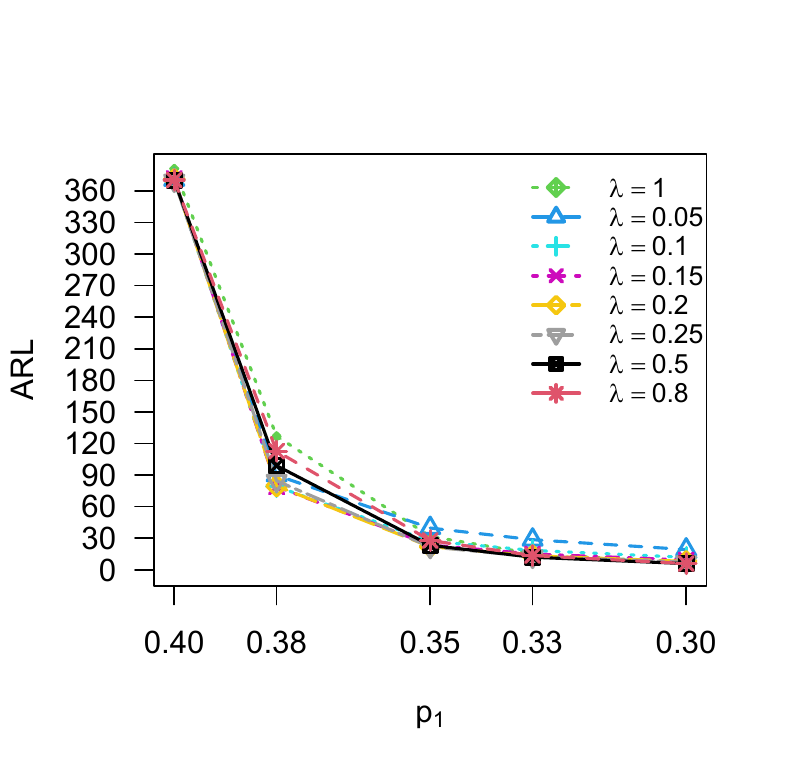}
		\caption{$k=5$}
		\label{3.2d}
	\end{subfigure}
	\caption{Performance of the ZINB EWMA and Shewhart charts with $n=20, \theta=0.85$, and $k=1, 2, 3, 5$ using $ARL_0 = 370$}
	\label{fig:2}
\end{figure}

Next, a graphical comparison of the ZINB-EWMA and ZINB-Shewhart charts is discussed. It is noticed from Figure 
\ref{fig:2}, the ZINB-EWMA performs better in terms of ARL with different values of $k$ and $\lambda$ than the ZINB-Shewhart chart. 

\section{Real data Application}
A real data set is used in this study to illustrate the application of the proposed chart.

\subsection{Owls Data Set}
The data set refers to owls dataset is taken from the R-package \textbf{''glmmTMB"} \citep{roulin2007nestling}. Roulin and Bersier\cite{2043} looked at how nestlings responded to the presence of the father and of the mother. Using microphones inside and a video outside the nests, they sampled 27 nests and studied vocal begging behaviour when the parents brought prey. The number of nestlings was between 2 and 7 per nest. The variables of the data set are: 
\begin{itemize}
	\item Nest a factor: describing individual nest, 
	\item Food Treatment (factor): Deprived or Satiated locations,  
	\item Sex Parent (factor) sex of provisioning parent: Female or Male, 
	\item ArrivalTime: a numeric vector, \item SiblingNegotiation: a numeric vector, \item BroodSize: brood size, \item NegPerChick: number of negotiations per chick
\end{itemize}
There are 599 observations in the data and for the monitoring purpose, the Sibling Negotiation, ranging from 100 to 350 are extracted from the data set. The sibling negotiation is defined as follows. Using the recorded footage, the number of calls made by all offspring in the absence of the parents was counted during 30-s time intervals every 15 min. To allocate a number of calls to a visit from a parent, the counted number of calls from the preceding 15 min of the arrival was used. This number was then divided by the number of nestlings. Thus, it is just the number of counted calls in the nearest 30-s interval before the arrival of a parent divided by the number of nestlings \citep{2044}.

\begin{figure}[H]
	\centering
	\begin{subfigure}[b]{0.48\textwidth}
		\centering
		\includegraphics[width=\textwidth]{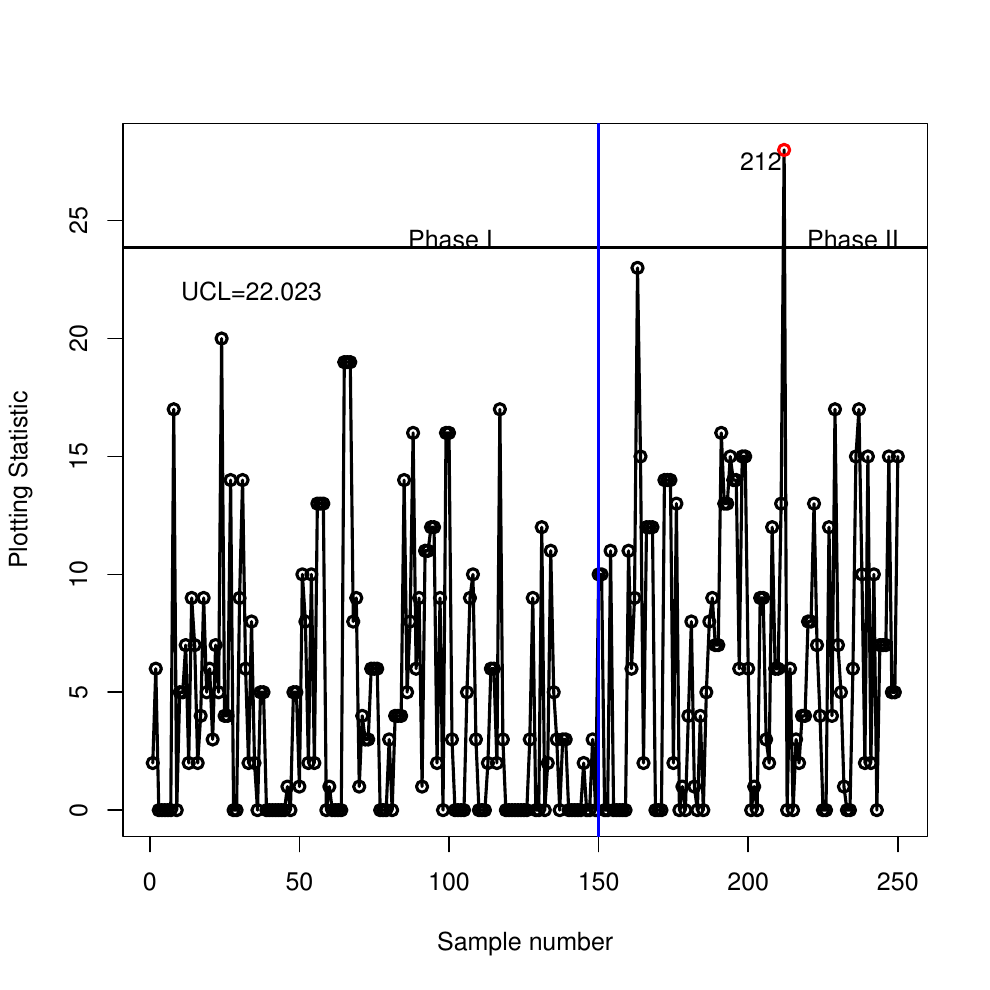}
		\caption{Shewhart}
	\label{fig:shewhart 2os}
	\end{subfigure}
	\hfill
	\begin{subfigure}[b]{0.48\textwidth}
		\centering
		\includegraphics[width=\textwidth]{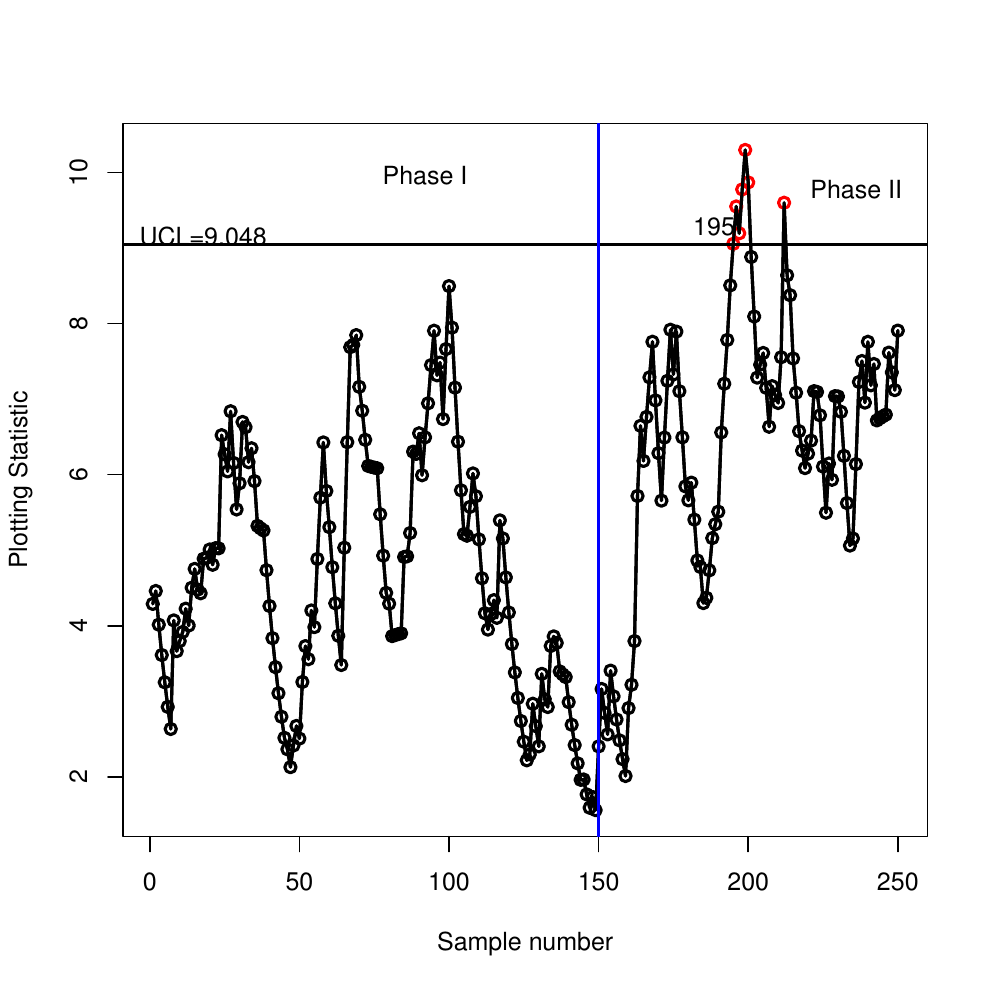}
		\caption{EWMA}
		\label{fig:ewma-2}
	\end{subfigure}
	\caption{Phase I and II ZINB-Shewhart and ZINB-EWMA control charts for Owls Data}
	\label{Realdata1}
\end{figure}

For the IC process, the first 150 observations are selected, while the rest of the observations are considered as the Phase-II data. The Bayesian information criterion (BIC) is used to select the best model among, NB, ZIP, Poisson, and ZINB. As listed in Table \ref{tab:3.73}, the ZINB has the lowest value of BIC and thus, it is the most suitable model to describe the data. Note that $\hat p=\frac{\hat k(1-\hat\theta)}{\hat\mu+\hat k(1-\hat\theta)}$.
\begin{table}[H]
	\centering
	\small
	\caption{Model fitting on Sibling Negotiation of Owls dataset}
	\begin{tabular}{c|c|c|c|c|c|c}
		\hline
		 Mean &  Lower $95\%$ CI& Upper $95\%$ CI & BIC & $\hat k$   & $\hat\theta$ & Model \\
		\hline
		  6.719 & 6.5151 & 6.9303 & 5978.11 &-  & -  & Poisson \\
		  8.823 & 8.2113 & 9.4817 & 3449.18 & 2.0 & 0.24 & ZINB \\
		 6.719 & 6.6073 & 7.4343 & 3551.22 & 1.0 & -  & NB \\ 
		 9.085 & 8.8082 & 9.3699 & 4242.85 & -  & 0.26 & ZIP \\
		\hline
	\end{tabular}%
	\label{tab:3.73}%
\end{table}%
We implemented EWMA and Shewhart chart and it is noticed that no point is OOC in the Phase I data. However, point 212 is OOC using the Shewhart chart (Figure \ref{fig:shewhart 2os}), while seven points 195, 196, 197, 198, 199, 200, and 212, are OOC using the EWMA chart (Figure \ref{fig:ewma-2}). 
\subsection{Dispersion Tests}
	\begin{table}[H]
		\centering
		\small
		\caption{Dispersion Tests}
		\begin{tabular}{cccccccc}
			\hline
			&  \multicolumn{3}{c}{Naive Test}& Dispersion Test 1& Dispersion Test2&Dispersion Para. &c \\
			Data &Mean& Var&CV&pval &pval & &\\
			\hline
			Owls &6.7195&44.4998&0.9928&$<0.0001$&$<0.0001$&0.6921&5.6114  \\
			
			\hline
		\end{tabular}%
		\label{tab:3.713}%
	\end{table}%
	For the real data sets, the most important thing is to decide whether data is over-dispersed or not, as pointed out by the editor. To this end, in the R package AER \citep{aer}, a function \textit{dispersiontest} can be used to implement a test for over-dispersion \citep{Cameron}. It follows a simple idea. In a Poisson model, we have $E(Y)=\mu=Var(Y)$. The test simply tests this assumption as a null hypothesis against an alternative where $Var(Y)=\mu+c*f(\mu)$, where the constant $c<0$ means under-dispersion and $c>0$ means over-dispersion. The function $f(.)$ is some monotonic function (often linear or quadratic; the former is the default). The resulting test is equivalent to testing $H_0:c=0$ versus $H_1:c\neq0$ and a t statistic is used which asymptotically follows standard normal under the null hypothesis. From Table \ref{tab:3.713}, one can clearly see that there is evidence of over-dispersion ($c$ is estimated to be 5.5 although the naive test by CV is close to one for the Owls data), which speaks quite strongly against the assumption of equi-dispersion (i.e., $c=0$). Also, the p-value (dispersion test 2) is less than a 5\% level of significance. 
	An alternative is the odTest from the pscl library\citep{pscl}, which compares the log-likelihood ratios of a negative binomial regression to the restriction of a Poisson regression $\mu=Var$. The p-values reported for the Owls data set in Table \ref{tab:3.713} is less than 5\% (dispersion test 1). Thus, the null hypothesis of the Poisson restriction is rejected in favour of negative binomial. The third test is to use a likelihood-ratio test to show that a quasipoisson generalized linear model (GLM) with over-dispersion is significantly better than a regular Poisson GLM without over-dispersion. Again, we obtained p-values less than 5\% level of significance. Thus, the data set is overdispersed.
\section{Conclusion}
In many cases, we deal with the count data, like COVID-19 cases, the number of goals in a football world cup match, etc. In such cases, the Poisson distribution is widely used. However, this distribution cannot deal with excessive zeros and over-dispersion, i.e., variance greater than the mean leads to over-dispersion. Generally, the ZINB distribution is preferred in such cases than the Poisson distribution. In comparison to the Poisson distribution, the ZINB distribution more effectively handles over-dispersion in a data set, due to its dispersion parameter $k$. 

This study proposed the ZINB-EWMA and ZINB-Shewhart control charts and assessed their performance using the ARL and SDRL measures. Different values of $\theta$, such as (0.85, 0.80, 0.75) and $k$ for the ARL computation of the ZINB charts are used. The results suggested that using $\theta=0.75$ with $k=3, 5$ and a large shift in $p$ like $p_1=0.33, 0.35$, the ZINB-Shewhart control chart performs better than the ZINB-EWMA with $\lambda=0.05$. For small shifts, the ZINB-EWMA performs well in all cases of smoothing parameter $\lambda$. If the value of $\theta$ is decreased in the ZINB-EWMA chart, the value of $L$ also decreases. The same is the case of ZINB-Shewhart control chart, i.e., the value of $L$ decreases when $\theta$ decreases. Both charts indicate that the value of $L$ decreased as the value of $k$ increased. The simulated result shows that ZINB-EWMA is more efficient than the ZINB-Shewhart control chart. In addition, the real data examples show the superiority of the EWMA chart over the Shewhart chart.

In the future, the research work can be extended to cumulative sum and adaptive control charts. Furthermore, variable sample size and variable sampling interval charts can be developed. Besides bivariate extension of the ZINB chart, the ARL-unbiased design can also be proposed.  

\section*{Acknowledgments}
The authors would like to thanks three anonymous referees, associate editor, and editor for the constructive suggestions to improve the quality and presentation of our work.
\bibliographystyle{vancouver}
\bibliography{ref_bib}

\begin{thebibliography}{10}

\bibitem{New8}
Saghir A, Lin Z.
\newblock Control charts for dispersed count data: an overview.
\newblock Quality and Reliability Engineering International. 2015;31(5):725-39.

\bibitem{New1}
Mahmood T, Xie M.
\newblock Models and monitoring of zero-inflated processes: {The} past and
  current trends.
\newblock Quality and Reliability Engineering International.
  2019;35(8):2540-57.

\bibitem{Chananet}
Chananet C, Sukparungsee S, Areepong Y.
\newblock The {ARL} of {EWMA} Chart for Monitoring {ZINB} Model Using {Markov
  Chain} Approach.
\newblock International Journal of Applied Physics and Mathematics.
  2014;4(4):236-9.

\bibitem{New5}
Katemee N, Mayureesawan T.
\newblock Control charts for zero-inflated {Poisson} models.
\newblock Applied Mathematical Sciences. 2012;6(56):2791-803.

\bibitem{New3}
Fatahi AA, Noorossana R, Dokouhaki P, Moghaddam BF.
\newblock Zero inflated {Poisson EWMA} control chart for monitoring rare
  health-related events.
\newblock Journal of Mechanics in Medicine and Biology. 2012;12(04):1250065.

\bibitem{New9}
Rakitzis AC, Castagliola P.
\newblock The effect of parameter estimation on the performance of one-sided
  {Shewhart} control charts for zero-inflated processes.
\newblock Communications in Statistics -Theory and Methods.
  2016;45(14):4194-214.

\bibitem{New2}
Mukherjee A, Rakitzis AC.
\newblock Some simultaneous progressive monitoring schemes for the two
  parameters of a zero-inflated {Poisson} process under unknown shifts.
\newblock Journal of Quality Technology. 2019;51(3):257-83.

\bibitem{New7}
Urbieta P, Lee~Ho L, Alencar A.
\newblock {CUSUM and EWMA} control charts for negative binomial distribution.
\newblock Quality and Reliability Engineering International.
  2017;33(4):793-801.

\bibitem{New10}
Lai X, Liu R, Liu L, Wang J, Zhang X, Zhu X, et~al.
\newblock Residuals based {EWMA} control charts with risk adjustments for
  zero-inflated Poisson models.
\newblock Quality and Reliability Engineering International.
  2022;38(1):283-303.

\bibitem{New13}
Hu Q, Liu L.
\newblock Weighted Score test based {EWMA} control charts for Zero-Inflated
  {Poisson} Models.
\newblock Computers \& Industrial Engineering. 2021;152:106966.

\bibitem{New14}
Asghar M, Ali S, Shah I.
\newblock Poisson hurdle model for monitoring the inflation of zeros.
\newblock Quality and Reliability Engineering International.
  2023;39(6):2152-61.

\bibitem{greene1994accounting}
Greene W.
\newblock Accounting for Excess Zeros and Sample Selection in {Poisson} and
  Negative Binomial Regression Models.
\newblock New York University, Leonard N. Stern School of Business, Department
  of Economics; 1994.
\newblock Available from:
  \url{https://EconPapers.repec.org/RePEc:ste:nystbu:94-10}.

\bibitem{sharma2013zero}
Sharma A, Landge V.
\newblock Zero inflated Negative binomial for modeling heavy vehicle crash rate
  on {Indian} rural highway.
\newblock International Journal of Advances in Engineering \& Technology.
  2013;5(2):292-301.

\bibitem{jansakul2008score}
Jansakul N, Hinde JP.
\newblock Score tests for extra-zero models in zero-inflated negative binomial
  models.
\newblock Communications in Statistics-Simulation and Computation.
  2008;38(1):92-108.

\bibitem{zulkifli2011zero}
Zulkifli M, Ismail N, Razali AM.
\newblock Zero-inflated {Poisson} versus zero-inflated negative binomial:
  Application to theft insurance data.
\newblock The 7th IMT-GT International Conference on Mathematics, Statistics
  and its Applications Thailand; 2011. .

\bibitem{aa2012analysis}
Aa MA, Naing NN.
\newblock Analysis death rate of age model with excess zeros using zero
  inflated negative binomial and negative binomial death rate: Mortality aids
  co-infection patients, {K}elantan {M}alaysia.
\newblock Procedia Economics and Finance. 2012;2:275-83.

\bibitem{saputro2021estimation}
Saputro MIA, Qudratullah MF.
\newblock Estimation of Zero-Inflated Negative Binomial Regression Parameters
  Using the Maximum Likelihood Method (Case Study: Factors Affecting Infant
  Mortality in {Wonogiri} in 2015).
\newblock In: Proceeding International Conference on Science and Engineering.
  vol.~4; 2021. p. 240-54.

\bibitem{cheung2002zero}
Cheung YB.
\newblock Zero-inflated models for regression analysis of count data: a study
  of growth and development.
\newblock Statistics in Medicine. 2002;21(10):1461-9.

\bibitem{rose2006use}
Rose CE, Martin SW, Wannemuehler KA, Plikaytis BD.
\newblock On the use of zero-inflated and hurdle models for modeling vaccine
  adverse event count data.
\newblock Journal of Biopharmaceutical Statistics. 2006;16(4):463-81.

\bibitem{hall2000zero}
Hall DB.
\newblock Zero-inflated {Poisson} and binomial regression with random effects:
  {A} case study.
\newblock Biometrics. 2000;56(4):1030-9.

\bibitem{yau2001zero}
Yau KK, Lee AH.
\newblock Zero-inflated {Poisson} regression with random effects to evaluate an
  occupational injury prevention programme.
\newblock Statistics in Medicine. 2001;20(19):2907-20.

\bibitem{hall2005two}
Hall DB, Wang L.
\newblock Two-component mixtures of generalized linear mixed effects models for
  cluster correlated data.
\newblock Statistical Modelling. 2005;5(1):21-37.

\bibitem{williamson2007power}
Williamson JM, Lin H, Lyles RH, Hightower AW.
\newblock Power calculations for {ZIP and ZINB} models.
\newblock Journal of Data Science. 2007;5(4):519-34.

\bibitem{hashim2021application}
Hashim L, Dreeb N, Hashim K, Shiker MA.
\newblock An application comparison of two negative binomial models on rainfall
  count data.
\newblock In: Journal of Physics: Conference Series. vol. 1818. IOP Publishing;
  2021. p. 012100.

\bibitem{adarabioyo2019comparing}
Adarabioyo M, Ipinyomi R.
\newblock Comparing zero-inflated {Poisson}, zero-inflated negative binomial
  and zero-inflated geometric in count data with excess zero.
\newblock Asian Journal of Probability and Statistics. 2019;4(2):1-10.

\bibitem{sheu2004effect}
Sheu Ml, Hu Tw, Keeler TE, Ong M, Sung HY.
\newblock The effect of a major cigarette price change on smoking behavior in
  {California}: a zero-inflated negative binomial model.
\newblock Health Economics. 2004;13(8):781-91.

\bibitem{New17}
Fang R. Zero-inflated negative binomial {(ZINB)} regression model for
  over-dispersed count data with excess zeros and repeated measures: {An}
  application to human microbiota sequence data [Master Thesis]; 2013.

\bibitem{mundform2011number}
Mundform DJ, Schaffer J, Kim MJ, Shaw D, Thongteeraparp A, Supawan P.
\newblock Number of replications required in {Monte Carlo} simulation studies:
  A synthesis of four studies.
\newblock Journal of Modern Applied Statistical Methods. 2011;10(1):4.

\bibitem{roulin2007nestling}
Roulin A, Bersier LF.
\newblock Nestling barn owls beg more intensely in the presence of their mother
  than in the presence of their father.
\newblock Animal Behaviour. 2007;74(4):1099-106.

\bibitem{2043}
Roulin A, Bersier LF.
\newblock Nestling barn owls beg more intensely in the presence of their mother
  than in the presence of their father.
\newblock Animal Behaviour. 2007;74(4):1099-106.

\bibitem{2044}
Zuur A, Ieno EN, Walker N, Saveliev AA, Smith GM.
\newblock Mixed Effects Models and Extensions in Ecology with {R}.
\newblock Statistics for Biology and Health. Springer New York; 2009.
\newblock Available from:
  \url{https://books.google.com.pk/books?id=vQUNprFZKHsC}.

\bibitem{aer}
Kleiber C, Zeileis A. AER: {Applied Econometrics with R}; 2022.
\newblock R package version 1.2-10.
\newblock Available from:
  \url{https://rdocumentation.org/packages/AER/versions/1.2-10}.

\bibitem{Cameron}
Cameron AC, Trivedi PK.
\newblock Regression-based Tests for Overdispersion in the {Poisson} Model.
\newblock Journal of Econometrics. 1990;46:347-64.

\bibitem{pscl}
Jackman S, Tahk A, Zeileis A, Maimone C, Fearon J, Meers Z. pscl: Political
  Science Computational Laboratory; 2022.
\newblock R package version 1.5.5.
\newblock Available from:
  \url{https://cran.r-project.org/web/packages/pscl/index.html}.

\bibitem{Abbas12102024}
Abbas A, Ali S, Shah I.
\newblock Exponentially weighted moving average chart using zero-inflated
  negative binomial distribution.
\newblock Journal of Statistical Computation and Simulation.
  2024;94(15):3375-90.

\end{thebibliography}
\section*{Corrigendum: Exponentially Weighted Moving Average Chart using Zero-Inflated Negative Binomial Distribution}
{\color{red}
The values in Table \ref{tab:4n} were mistakenly replaced in the originally published article Abbas et al.\cite{Abbas12102024}. The correct table and its interpretation are provided below.

\renewcommand{\thetable}{13}
\begin{table}
	\centering
	\small
	\caption{ZINB-EWMA ($\lambda=0.10$) and ZINB-Shewhart ($\lambda=1.0$) for $n=10$ assuming shifts in all parameters and $p_0=0.4,\theta_0=0.85,k_0=1$}
	\begin{tabular}{|ccc|cc|cc|cc|}
		\toprule
		& &&\multicolumn{2}{c|}{Shewhart, L=4.891 ($ARL_0=500$)} & \multicolumn{2}{c|}{EWMA L=3.083 ($ARL_0=500$)} & \multicolumn{2}{c|}{EWMA L=3.012 ($ARL_0=449$)} \\
		\hline
		$p_1$ & $\theta_1$ & $k_1$   & ARL & SDLR & ARL & SDLR & ARL & SDLR \\
		\hline
		{\bf0.4} & {\bf0.85} & {\bf1}   & {\bf449.98} & 455.44 & {\bf500.89} & 497.11 & {\bf449.8} & 445.58 \\
		&     & 2   &38.13  &37.61  &16.14  &11.76  &15.60  &11.39  \\
		&     & 3   &11.56  &10.94  &6.97  &4.43  &6.81  &4.34  \\
		&     & 5   &3.80  &3.22  &3.55  &2.14  &3.49  &2.10  \\
		0.4 & 0.8 & 1   &222.97  &223.36  &83.07  &77.35  &77.15  & 72.10 \\
		&     & 2   &19.93  &19.48  &8.60  &5.27  &8.36  &5.12  \\
		&     & 3  &6.55  &5.99  &4.52  &2.55  &4.43  &2.49  \\
		&     & 5  &2.52  &1.95  &2.56  &1.41  &2.53  &1.39  \\
		0.4 & 0.75 & 1   &124.85  &125.06  &30.93  &24.91  &29.41  &23.70  \\
		&     & 2   &12.06  &11.66  &5.83  &3.18  &5.71  &3.13  \\
		&     & 3   &4.28  &3.68  &3.39  &1.75  & 3.34 &1.72  \\
		&     & 5   &1.9  &1.3  &2.04  &1.03  &2.03  &1.01  \\
		0.38 & 0.85 & 1   &290.22  &286.90  &244.25  &239.46  &223.16  &218.14  \\
		&     & 2   &28.04  &27.49  &13.09  &9.34  &12.70  &9.05  \\
		&     & 3   &9.05  &8.53  &6.11  &3.84  &5.99  &3.78  \\
		&     & 5  &3.30  &2.75  &3.23  &1.97  &3.19  &1.93  \\
		0.38 & 0.8 & 1   &146.03  &146.43  &52.41  &46.11  &49.32  &43.37  \\
		&     & 2   &15.05  &14.27  &7.36  &4.43  &7.17  &4.33  \\
		&     & 3   &5.35  &4.78  &4.08  & 2.31 &4.01  &2.26  \\
		&     & 5   &2.27  &1.71  &2.37  &1.31  &2.36  &1.29  \\
		0.38 & 0.75 & 1   &84.22  &82.99  &22.49  &16.98  &21.52  &16.15  \\
		&     & 2   &9.33  &8.76  &5.14  &2.78  &5.03  &2.73  \\
		&     & 3  &3.60  &3.06  &3.08  &1.60  &3.04  &1.58  \\
		&     & 5  &1.75  &1.15  &1.91  &0.96  &1.90  &0.95  \\
		{0.35} & {0.85} & 1   &151.83  &150.21  &99.79  &95.79  &92.31  &88.02  \\
		&     & 2   &18.16  &17.78  &9.92  &6.80  &9.64  &6.59  \\
		&     & 3   &6.61  &6.06  &5.08  &3.18  &4.98  &3.11  \\
		&     & 5   &2.80  &2.25  &2.85  &1.75  &2.82  &1.70  \\
		{0.35} & {0.8} & 1   &80.06  &80.28  &29.68  &24.28  &28.36  &23.12  \\
		&     & 2   &10.14  &9.53  &5.96  &3.54  &5.85  &3.48  \\
		&     & 3   &4.11  &3.58  &3.51  &2.00  &3.45  &1.96  \\
		&     & 5   &1.98  &1.40  &2.13  &1.18  &2.11  &1.16  \\
		{0.35} & {0.75} & 1   &46.59  &46.44  &15.42  &10.72  &14.86  &10.31  \\
		&     & 2   &6.52  &5.94  &4.32  &2.32  &4.24  &2.28  \\
		&     & 3   &2.86  &2.30  &2.71  &1.41  &2.68  &1.38  \\
		&     & 5  &1.58  &0.97  &1.74  &0.87  &1.74  &0.86  \\
		{0.33} & {0.85} & 1   &103.28  &101.80  &61.06  &56.04  &57.56  &52.56  \\
		&     & 2   &13.84  &13.36  &8.38  &5.60  &8.15  &5.43  \\
		&     & 3   &5.45  &4.85  &4.54  &2.84  &4.46  &2.78  \\
		&     & 5   & 2.52 &1.98  &2.64  &1.62  &2.61  &1.59  \\
		{0.33} & {0.8} & 1   &54.39  &54.76  &21.72  &16.88  &20.86  &16.08  \\
		&     & 2   &7.98  &7.36  &5.26  &3.08  &5.15  &3.02  \\
		&     & 3  &3.49  &2.94  &3.20  &1.82  &3.16  &1.77  \\
		&     & 5   &1.84  &1.26  &1.99  &1.11  &1.99  &1.09  \\
		{0.33} & {0.75} & 1   &33.05  &32.89  &12.26  &8.06  &11.88  &7.82  \\
		&     & 2  &5.23  &4.61  &3.85  &2.08  &3.79  &2.04  \\
		&     & 3   &2.51  &1.94  &2.50  &1.29  &2.48  &1.27  \\
		&     & 5   &1.48  &0.85  &1.65  &0.81  &1.64  &0.80  \\
		\bottomrule
	\end{tabular}%
	\label{tab:4n}%
\end{table}%

In the previous tables, we studied the performance of the charts assuming shifts in the parameter $p$. In practice, a shift can appear in any parameter of the ZINB distribution. To show the performance for this case, Table \ref{tab:4n} show the ARL and SDRL values of the ZINB-EWMA and the ZINB-Shewhart for $n=10$ with $ARL_0=500$ (fourth and sixth columns) and $ARL_0=449$ (eighth column to match with the $ARL_0$ achieved by the Shewhart chart in the fourth column). Table \ref{tab:4n} tabulates the results for mean monitoring using Shewhart and EWMA charts. Again, it is noticed that the $ARL_0$ is not achieved for the Shewhart chart. The $ARL_1$ for EWMA and Shewhart charts decreased as $p_0$ shifted to $p_1$. However, the EWMA chart outperform the Shewhart chart. In particular, the IC ARL of the ZINB-EWMA chart with parameters (n, k, $p_0$, $\theta$)=(10, 1, 0.40, 0.85), $\lambda$=0.10, and L=3.083 is $ARL_0$=500.89 with $SDRL_0$=497.11. For the Shewhart chart assuming (n, k, $p_0$, $\theta$)=(10, 1, 0.40, 0.85) with $\lambda$=1 and L=4.891, the IC process will give the $ARL_0$=449.98 with $SDRL_0$=455.44. For the OOC specifications (n, k, $p_1$, $\theta$)=(10, 1, 0.40, 0.80) and $p_1$=0.40, the $ARL_1$ and $SDRL_1$ for the ZINB-EWMA chart with $\lambda$=0.1 and L=3.083 are 83.07 and 77.35, respectively. However, for (n, k, $p_1$, $\theta$)=(10, 1, 0.40, 0.75) and $p_1$=0.38, the $ARL_1$ for the ZINB-EWMA chart with $\lambda$=0.10 and L=3.083 are 30.93 and 244.25, respectively. For the ZINB-Shewhart chart with (n, k, $p_1$, $\theta$)=(10, 2, 0.40, 0.85), $\lambda$=1, and L=4.891, $ARL_1=38.13$ and $SDRL_1=37.61$. Similarly, when $p_1$=0.38 and (n, k, $p_0$, $\theta$)=(10, 1, 0.40, 0.85), the $ARL_1$ and $SDRL_1$ of the ZINB-Shewhart with $\lambda$=1 and L=4.891 are 290.22 and 286.90, respectively. Generally, it is noticed that the ARL values of both charts decrease as the values of $k$ increased. The ZINB-EWMA performs well as compared to the ZINB-Shewhart, because the ZINB-EWMA give small value of $ARL_1$ in all cases of $p$ and $k$.

The last two columns of Table \ref{tab:4n} list the ARL and SDRL values of the ZINB-EWMA for $n=10$ assuming $ARL_0=449$ as this IC ARL is achieved in the 3rd column of the same table. It is again evident that the EWMA chart outperforms the Shewhart chart. }
\end{document}